# General linear correction method for DFT+$X$ energy: application to U-M (M=Al, Ga, In) alloys under high pressure

Xiao L. Pan[1,2,3], Hong X. Song[1], Y. Sun[1], F. C. Wu[1]*, H. Wang[1], Y. F. Wang[1], Y. Chen[5], Xiang R. Chen[2]*, Hua Y. Geng[1,4]*

[1] *National Key Laboratory of Shock Wave and Detonation Physics, Institute of Fluid Physics, China Academy of Engineering Physics, Mianyang, Sichuan 621900, P. R. China;*

[2] *College of Physics, Sichuan University, Chengdu 610065, P. R. China;*

[3] *Extreme Condition Material Properties Science and Technology center, School of Mathematics and Physics, Southwest University of Science and Technology, Mianyang, 621010, P. R. China;*

[4] *HEDPS, Center for Applied Physics and Technology, and College of Engineering, Peking University, Beijing 100871, P. R. China;*

[5] *Fracture and Reliability Research Institute, School of Engineering, Tohoku University, Sendai 980-8579, Japan.*

**Abstract:** DFT+$X$ methods, such as DFT+$U$ and DFT+DMFT, are important supplements to standard density functional theory when strong on-site Coulomb interactions are present. However, the involvement of external parameters in the underlying model Hamiltonian introduces intrinsic ambiguity when comparing the total energies obtained with different model parameters. This renders DFT+$X$ approaches semi-empirical and severely hinders their capability to describe phase ordering and phase stability, especially when reliable experimental benchmarks are unavailable, such as under high pressure. In this work, we resolve this longstanding problem by proposing a general linear correction method that eliminates the ambiguous energy contributions introduced by the model Hamiltonian in DFT+$X$ approaches, thereby enabling direct comparison of their energies calculated with different interaction parameters. The method is demonstrated and validated within the framework of DFT+$U$, an important member of the DFT+$X$ family. It is then applied to important nuclear materials of uranium-based binaries U-M (M=Al, Ga, In) alloys. With this approach, we resolve the long-standing discrepancy between theoretical predictions and experimental observations of phase stability with unprecedented accuracy, and predict several previously unknown stable intermetallic compounds under high pressure. The broad applicability of the method is further confirmed by accurate predictions of formation enthalpies for diverse systems, including Np-Al, U-Si, and Cu-O binaries, the ternary MnSnAu compound, and oxygen adsorption on the Cu(111) surface. This work establishes linear-corrected DFT+*U* as a fully first-principles approach and validates the linear correction method as a robust and general scheme that can be readily extended to other DFT+*X* methods.

* *Corresponding authors. E-mail: s102genghy@caep.cn; xrchen@scu.edu.cn; fcwu2011@mail.ustc.edu.cn*

# I. Introduction

Due to its good balance between high efficiency and accuracy, density functional theory (DFT) [1, 2] has grown into a workhorse of condensed matter physics and materials science [3-6], where phase stability/ordering and phase diagrams are usually among the most important physical properties that are widely investigated. Regardless of its initial success in simple metals, DFT suffers from the self-interaction error (SIE) [7] and provides an inaccurate description of strong electronic correlations when orbitals become localized, a deficiency intrinsic to all static mean-field theories. This causes the commonly used local and semi-local exchange–correlation functionals, such as the local density approximation (LDA) [2]， the generalized gradient approximation (GGA) [8], and even the most advanced SCAN functional [9], to underestimate band gaps in semiconductors and insulators [10, 11]. The total energies of strongly correlated systems with localized *d* or *f* electrons are also poorly described [12, 13]. Hybrid functionals, which incorporate a fraction of exact exchange, can partially alleviate this problem; however, they significantly increase the computational cost and may introduce additional issues such as localization errors [14].

In contrast, combining DFT with the phenomenological Hubbard model for localized orbitals shows great promise in addressing these problems. This line of strategy has produced a family of methods collectively referred to as DFT+$X$, including DFT+$U$ [11, 15], DFT+G [16, 17], DFT+DMFT [18, 19], and DFT+$J$ [14]. For all of these methods, the model Hamiltonian depends on the empirical Hubbard parameters $U$ and $J$ (or the effective $U$-$J$), which describe the strength of on-site Coulomb interactions. The total energy of DFT+$X$ is thus a function of these external parameters, which are unfortunately not well-defined and contain ambiguity. The most notorious consequence is that energies calculated with different values of these parameters cannot be directly compared. Even worse, these parameters usually vary with pressure [12], chemical compositions, and even local atomic environment [20], making the determination of phase stability and phase diagrams using DFT+$X$ methods highly problematic, if not impossible. To address this issue, several methods that rely on experimental information have been proposed in recent years [21, 22, 23, 24], primarily focusing on the DFT+$U$ formalism because of its simplicity and high computational efficiency. The extension of these methods to other DFT+$X$ approaches is straightforward.

It should also be mentioned that some DFT+$U$ calculations in the literature directly compare uncorrected DFT+$U$ total energies obtained with different values of $U$. Such a practice, however, is very risky and should be adopted only when reliable experimental or theoretical data, such as cohesive energy or formation energy [21], are available and can be used as the reference. Similarly, for reactions in which only some reactants possess localized orbitals, calculating the reaction enthalpy requires comparing and mixing the energy obtained from DFT and DFT+$U$. In this situation, the plain approach mentioned above is clearly not applicable. To this end, Jain et al. proposed a mixed GGA/GGA+$U$ scheme [22] in which the DFT+$U$ energies are recalibrated with respect to available experimental data. Specifically, the GGA+$U$ energies are *adjusted* to align with the averaged experimental trends. This tentatively achieves the compatibility between the calculated DFT+$U$ and DFT energies, and yielding acceptable formation enthalpies. Another closely related method, the fitted elemental-phase reference energies (FERE) method [23, 24], uses the experimental formation enthalpies of selected compounds together with their DFT+$U$ total energies to derive the *artificial* elemental reference phase energies by solving a linear least-squares problem. These reference energies are then employed to compute the formation enthalpies of other compounds that are chemically and physically similar. Using this method, accurate formation enthalpies have been obtained for transition metal binary oxides [25] and metal sulfides [26], as intended.

Nevertheless, all of these methods rely entirely on experimental formation enthalpies to correct DFT+$U$ energies, rendering the latter essentially empirical. Unfortunately, accurate experimental data are scarce and often unavailable for most newly discovered materials. The same is true for common materials under extreme conditions, such as high pressure. Therefore, these methods (and their variants) cannot be applied in situations where no experimental data are available. In this context, accelerating materials design and discovery calls for a novel theoretical framework capable of correcting DFT+*X* energies without relying on experimental data.

In this work, we propose a general correction method for DFT+*X* energies (and enthalpies) that does not require any experimental or other theoretical reference data. The implementation of the method within the DFT+*U* framework is illustrated. The method is validated for U-Ga and U-Al systems at ambient conditions by comparison with available experimental data. It is then applied to U-M (M = Al, Ga, In) systems under high pressure, all of which are important nuclear materials. Several previously unknown phases have been predicted, which would have been inaccessible without the consistent theoretical framework provided by this correction method for accurately calculating formation enthalpies of strongly correlated systems using DFT+*X* under extreme conditions. Furthermore, the broad applicability of the method is demonstrated through successful applications to other complex systems, including binary systems such as U-Si, Np-Al, and Cu-O, as well as the adsorption formation enthalpy of oxygen on the Cu(111) surface and the formation enthalpy of the ternary MnSnAu compound.

## II. Theoretical formalism

### A. General consideration

It is evident that the goal of all the aforementioned corrections to the DFT+*U* energies is to render them compatible with pure DFT, such that the resulting energies can be directly compared. Previously, this is done by removing the ambiguity in DFT+*U* energy by directly using experimental formation enthalpies in the mixed GGA/GGA+*U* method [22], or by introducing artificial elemental reference energies with the aid of experimental data in the FERE method [24]. Nevertheless, we should point out that it is theoretically possible to remove such ambiguity without resorting to any experimental guidance. This fully *ab initio* capability is highly desirable, particularly given the rapidly expanding classes of novel materials and their applications under extreme conditions, where reliable energy information is essential but experimental data are often unavailable. Unfortunately, no such approach has been reported in the literature to date.

Such a correction may be formulated at the level of local atomic approximation, namely by assuming that the unphysical contribution responsible for the ambiguity is confined within individual atoms. We are reminded of the cohesive energy definition of condensed matter, which ensures compatibility among different exchange–correlation (XC) functionals in DFT. Following similar idea, a "cohesive energy" can be defined within the DFT+*U* framework by subtracting the atomic energy—calculated using DFT+*U* with the same value of the Hubbard parameters—from the total energy. A practical difficulty of this approach is that, when performing DFT+*U* calculations for isolated atoms, most currently available DFT codes encounter severe numerical convergence problems. This technical limitation prevents the application of such a naive local atomic correction.

On the other hand, we note that in all DFT+*X* methods the introduced Hubbard model Hamiltonian contains terms that depend linearly on the model parameters such as *U* and/or *J*. These extrinsic terms have two physical effects: (*i*) they tune the electronic density to a more appropriate distribution, (*ii*) they contribute to the energy directly. According to the Hohenberg-Kohn theorem [1], the ground-state electronic density uniquely determines the entire electronic system. Therefore, if a DFT+*X* method properly describes a system, it should produce an accurate density, and this density must be close to that of the exact DFT. Although there is no exact XC functional for DFT available yet, advanced functionals such as SCAN [9] can be employed to evaluate accurate energies from a given DFT+*X* density. Following this reasoning, we can combine the DFT+*U* density with the SCAN functional to remove the ambiguity in the former and make it compatible with the pure DFT. The theoretical basis of this approach is that the main error in DFT is the density-driven error and functional-driven error; and for a given density and a sufficiently accurate functional such as SCAN, the functional-driven error should be minor. In practice, one could first perform a DFT+*U* calculation to obtain the electronic density, which is then fed into the pure DFT with SCAN functional to acquire the energy while keeping the density unchanged. This non-self-consistent scheme, which combines the DFT+*U* density with the SCAN functional, is inspired by previous work aimed at correcting density-driven errors in pure DFT such as Ref. [27], and is conceptually similar in nature. Unfortunately, in practice we found that this approach is applicable only at ambient conditions, where pressure or stress effects are negligible. Otherwise, this non-self-consistent approach usually gives a quite different pressure or stress from that of DFT+*U*, making it very difficult to align the enthalpy to the given pressure/stress, which is essential for calculating the phase stability and phase diagrams at elevated pressures.

**B. Linear correction method (LCM)**

Based on the general considerations as mentioned above, we introduce a linear correction method (LCM) to address this issue. Taking the simplified approach to DFT+*U* proposed by Dudarev *et al.* [11] as an example, the introduced Hubbard model Hamiltonian can be reduced to an energy functional supplemented to the pure DFT as：

$$E_{\mathrm{DFT}+U} = E_{\mathrm{DFT}} + \frac{U}{2}\sum_{\sigma}(n_{m,\sigma} - n_{m,\sigma}^{2}), \tag{1}$$

where *U* is the effective Hubbard parameter, $n_{m,\sigma}$ is the occupation number of the *m*-th state with a spin of $\sigma$. The resulting total energy expressed in terms of the Kohn-Sham eigenvalues is

$$E_{\mathrm{DFT}+U} = E_{\mathrm{DFT}}[\{\varepsilon_i\}] + \frac{U}{2}\sum_{l,j,\sigma}\rho_{lj}^{\sigma}\rho_{jl}^{\sigma}, \tag{2}$$

in which the second term represents the double-counting correction, and $\rho_{lj}^{\sigma}$ is the density matrix [11]. It is evident that the energy is a function of *U* and a functional of density, *i.e.*, *E*=*E* (*U*, [*n*]). In contrast, what is expected from DFT is purely *E*=*E*([*n*]) [1, 2]. The explicit dependence of the energy on *U* is unphysical and should be removed in order to obtain an unambiguous energy that can be directly compared with pure DFT results or experimental data. As can be seen from Eqs. (1) and (2), this dependence is *linear* with respect to $U$. Motivated by this observation, a linear correction method (LCM) for DFT+$U$ energy can be proposed as

$$E_{\mathrm{DFT}+U}^{\mathrm{LC}} = E_{\mathrm{DFT}+U} - \left(\frac{\partial E}{\partial U}\right)_{[n_{\mathrm{loc}}]} U \,. \tag{3}$$

Here, the superscript LC indicates the quantity already corrected by LCM, and the subscript $n_{\mathrm{loc}}$ in the second term indicates that the partial derivative is performed under the constraint that the localized charge density around the target atoms remains unchanged (Namely, in the sense of *local atomic approximation*). In practice, we can assume that for given systems that are chemically similar, the derivative in Eq. (3) can be approximated as a constant. As a result, the LCM corrected total energy per unit cell is given by

$$E_{\mathrm{DFT}+U}^{\mathrm{LC}} = E_{\mathrm{DFT}+U} - \epsilon \sum_{i=1}^{N} U_i \tag{4}$$

where $\epsilon$ is a constant, $U_i$ is the value of parameter $U$ for $i$-th atom, and $N$ is the total number of atoms in the unit cell. All quantities in Eq. (4) can be obtained directly from first-principles calculations.

To illustrate how the LCM can be applied to DFT+$X$, we consider a binary system as an example. Specifically, for a binary system $A_xB_y$, the formation enthalpy is defined as:

$$\Delta H_{A_xB_y}^{f} = (H_{A_xB_y} - xH_A - yH_B)/(x+y) \tag{5}$$

where $H_{A_xB_y}$ is the enthalpy per formula unit of the compound $A_xB_y$, and $H_A$ and $H_B$ represent the enthalpy per atom of elements A and B, respectively. At a given pressure, the enthalpy is given by $H=E+PV$, where $V$ represents the specific volume.

We assume that the localized orbitals of atom A are strongly correlated and therefore require special treatment within a DFT+$X$ framework (represented here by DFT+$U$ for simplicity), whereas atom B is weakly correlated and can be well described by pure DFT. We further note that the correction to $PV$ term is typically small. Therefore, the LCM is applied only to the internal energy $E$. This argument is justified because pressure is given by the first derivative of the energy with respect to volume. Within the framework of DFT+$U$, it can be expressed as

$$P = -\frac{\partial E([n],U)}{\partial V} = -\frac{\partial E}{\partial n}\Big|_U \frac{\partial n}{\partial V} - \frac{\partial E}{\partial U}\Big|_{[n]} \frac{\partial U}{\partial V}. \quad (6)$$

The variation of localized part $\frac{\partial E}{\partial U}\Big|_{[n_{\mathrm{loc}}]}$ is unphysical, and must be removed. We thus have

$$P_{\mathrm{DFT}+U}^{\mathrm{LC}} = P_{\mathrm{DFT}+U} + \epsilon \frac{\partial U}{\partial V}. \quad (7)$$

The second term in Eq. (7) is the contribution that requires LCM correction when evaluating the pressure. In general, the value of $U$ remains nearly constant over a small range of density variations, so that $\frac{\partial U}{\partial V} \approx 0$. However, for sufficiently large volume changes, the slope $\frac{\partial U}{\partial V}$ may become finite, and the pressure should then be corrected accordingly. However, if the magnitude of this correction is small, when we realign the enthalpy to the given pressure, the correction to enthalpy from the $PV$ term is cancelled by the leading correction from the internal energy, leaving only higher-order corrections that can be safely neglected [28]. Therefore, in any case, the correction to enthalpy can be approximated by the correction to the internal energy.

To apply LCM, we first notice from Eq. (4), that the energy difference between two chemically similar but compositionally different compounds (e.g., $A_xB_y$ and $A_{x'}B_{y'}$), after properly subtracting the contributions of the B atoms, is proportional to the constant $\epsilon$. That is,

$$\Delta E = (y'E^{\mathrm{DFT}+U}_{\mathrm{A}_x\mathrm{B}_y} - yE^{\mathrm{DFT}+U}_{\mathrm{A}_{x'}\mathrm{B}_{y'}}) - (y'E^{\mathrm{DFT}}_{\mathrm{A}_x\mathrm{B}_y} - yE^{\mathrm{DFT}}_{\mathrm{A}_{x'}\mathrm{B}_{y'}}) = \Delta N_U \cdot \epsilon \tag{8}$$

where $\epsilon$ is the constant defined in Eq. (4), and $\Delta N_U$ is defined as the difference in the effective Hubbard $U$ strength between the two compounds in the DFT+$U$ calculations, which is given by

$$\Delta N_U = y' N_U(\mathrm{A}_x\mathrm{B}_y) - yN_U(\mathrm{A}_{x'}\mathrm{B}_{y'}), \tag{9}$$

with

$$N_U(\mathrm{A}_x\mathrm{B}_y) = \sum_{i=1}^{N_A} U_i, \tag{10}$$

where $N_A$ is the total number of A-type atoms, and $U_i$ is the Hubbard $U$ value for the $i$-th A atom. For example, for compound $U_2Al_3$ in which each uranium atom has been described by DFT+$U$ with an effective Hubbard value of $U$= 2.5 eV, then its $N_U$ can be calculated as $N_U(\mathrm{U_2Al_3}) = \sum_{i=1}^{2} U_i$=2.5+2.5=5 eV.

For any given pair of compounds, once the $U$ values have been specified for each atom, the $\Delta N_U$ can be determined by Eq. (9), which then can be subsequently substituted into Eq. (8). On the other hand, the energy difference $\Delta E$ defined in Eq. (8) (*i.e.*, the first equality in Eq.(8)) can be obtained directly from the corresponding DFT+$U$ and DFT total energy calculations. The two quantities therefore can form a data pair ($\Delta E$, $\Delta N_U$). This procedure can be repeated for all possible pairs of compounds, yielding a set of ($\Delta E$, $\Delta N_U$) data points. The total number of such data pairs is $C_M^2$, where $M$ is the number of distinct compounds considered. For example, if we consider 10 different compositions of $A_xB_y$ compounds, a total of $C_{10}^2 = 45$ data points of ($\Delta E$, $\Delta N_U$) pair can be obtained. This data set can be used to assess the validity of the linear relationship as assumed in the second equality of Eq. (8). Moreover, it can be employed to determine the constant $\epsilon$ in Eq. (8) through linear fitting.

It is necessary to point out that if $\epsilon$ is not a constant, Eq. (8) cannot be held any more, and we need go back to Eq. (3). In any case, $\epsilon$ always can be expressed as a function of $U$, i.e., $\epsilon = \epsilon(U)$. Therefore, Eq. (3) should be replaced by

$$E_{\mathrm{DFT}+U}^{\mathrm{LC}} = E_{\mathrm{DFT}+U} - \int_0^U \epsilon(U') \cdot \mathrm{d}U', \tag{11}$$

in which $\epsilon(U') = \frac{\partial E}{\partial U}|_{[n_{\mathrm{loc}}],U=U'}$ and can be determined by the slope of $E([n],U)$ with respect to $U$ at the condition that the localized charge density $n_{\mathrm{loc}}$ is fixed, then the LCM is still applicable. As will be seen, in all cases we explored here, the linearity relation holds very well, and $\int_0^U \epsilon(U')dU'$ reduces to $\epsilon \cdot U$, which then gives Eq. (8).

Once Eq. (8) is validated and the value of $\epsilon$ constant is determined, the DFT+$U$ internal energy for each binary compound of $A_xB_y$ can be reliably corrected by

$$E_{\mathrm{DFT}+U}^{\mathrm{LC}} = E_{\mathrm{DFT}+U} - \Delta_U, \tag{12}$$

where $\Delta_U = \epsilon \cdot \sum_{i=1}^{N_A} U_i$. Similarly, the enthalpy is corrected by $H_{\mathrm{DFT}+U}^{\mathrm{LC}} = H_{\mathrm{DFT}+U} - \Delta_U$.

The formation enthalpy defined in Eq. (5) is then corrected by LCM as:

$$\Delta H_{\mathrm{A}_x\mathrm{B}_y}^{f,\mathrm{LC}} = \frac{(H_{\mathrm{A}_x\mathrm{B}_y}^{\mathrm{DFT}+U} - \Delta_U) - (xH_{\mathrm{A}}^{\mathrm{DFT}} + yH_{\mathrm{B}}^{\mathrm{DFT}})}{x+y}, \tag{13}$$

which can now be directly compared with experimental data and subsequently used to determine the phase stability and phase diagram of the system (here we have assumed that the elemental phase A and B do not require DFT+$U$ treatment).

Although DFT+$U$ is employed here as an illustrative example to demonstrate the LCM, the method itself does not depend on the specific theoretical details of how the Hubbard model is implemented. The same reasoning that leads to Eqs. (4, 7, 8, 12, 13) can be readily extended to other DFT+$X$ approaches. Such generalizations are straightforward and will not be elaborated here. It should also be pointed out that the LCM can be applied to multi-component systems containing two or more distinct strongly correlated species. The only thing that one should be cautioned is that each species could have its own constant $\epsilon$ in that situation.

## III. Validation and application

### A. Validation of LCM

We validate the LCM from two complementary aspects: First, we demonstrate that the linear relationship assumed in Eq. (8) holds in practice. Second, we compare the formation enthalpies obtained from Eq. (13) with available experimental data to show that accurate formation enthalpies can be achieved within DFT+$U$ without using any experimental information as input. For this purpose, we take Uranium-Gallium (U-Ga) system as a representative example. First-principles structure prediction methods are employed to identify the most stable candidate structures at zero pressure for ten possible stoichiometries of $U_xGa_y$ compounds (with $x$:$y$ = 4:1, 3:1, 2:1, 3:2, 1:1, 2:3, 3:5, 1:2, 1:3, 1:4). For these ten distinct compounds, there are $C_{10}^{2}=45$ unique compound pairs. Using Eqs. (8) and (9), we obtain 45 respective data points in the form of ($\Delta E$, $\Delta N_U$). The specific $U$ values adopted for uranium atoms in each compound are provided in Fig. S1 of the Supplementary Material (SM). These data are plotted in Fig. 1(a). As shown in Fig. 1(a), the data exhibit an excellent linear relationship, thereby validating the linearity assumption of Eq. (8). By performing a least-squares fit of Eq. (8) to the 45 first-principles data points, the dimensionless constant $\epsilon$ for the U-Ga system under ambient conditions is determined to be $\epsilon$=1.1122.

To assess the accuracy of LCM, we calculate the formation enthalpies using Eq. (13) with this $\epsilon$ value. The resulting convex hull constructed from the LCM-corrected formation enthalpies is shown by the red lines in Fig. 1(b), together with the pure DFT convex hull and the experimental data for U-Ga system [29, 30] for comparison. Note that these experimental data were measured at room temperature where the $U_xGa_y$ compounds are in non-magnetic (NM) state. To ensure a consistent comparison, all compounds are treated as NM in both the DFT+$U$ and pure DFT calculations. The uncorrected convex hull obtained directly from DFT+$U$ is provided in Fig. S2 of the SM. We find that the convex hull predicted by plain DFT+$U$ is unphysical: the formation enthalpies deviate significantly from the experimental data, and no stable compounds are predicted. Pure DFT performs better in this regard, with the shape of its convex hull qualitatively similar to the experimental one. However, the formation enthalpy of $UGa_3$ is overestimated by approximately 63.7% compared with the averaged experimental value. Moreover, pure DFT fails to predict the stability of $UGa_2$ and incorrectly yields two stable compounds, $U_2Ga$ and $U_2Ga_3$, indicating that pure DFT is insufficient to accurately describe the U–Ga system. In contrast, the LCM-corrected DFT+$U$ results yield an accurate convex hull for the U-Ga system and correctly predict the stability of $UGa_2$ and $U_2Ga_3$ at zero pressure. Notably, the predicted formation enthalpies are well bracketed by the experimental data: the predicted $\Delta H^{f,\mathrm{LCM}}_{\mathrm{UGa_3}}$ is just about 5.2% away from the calorimetry data, while that of $UGa_2$ is only 4.6% away from the average of the experimental data (about 10 folds better than pure DFT). Furthermore, the predicted crystal structures of $UGa_2$ (space group *P6/mmm*) and $UGa_3$ (space group *Pm-3m*) are in excellent agreement with experimental observations.

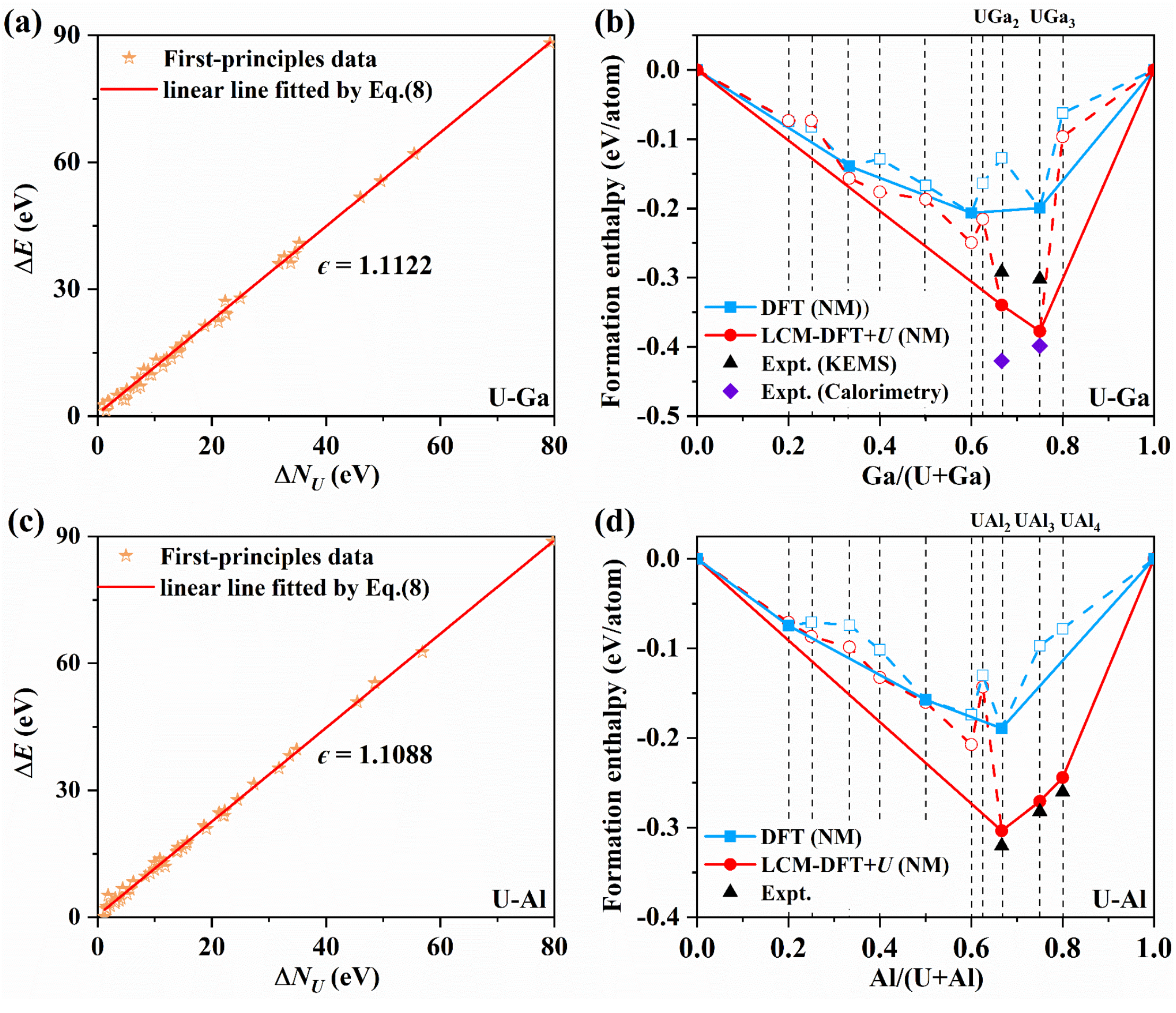

**Fig. 1** (color online) Calculated $\Delta E$-$\Delta N_U$ points and the corresponding linear fits obtained using Eq. (8) at ambient conditions for (a) U-Ga and (c) U-Al systems. Thermodynamic convex hulls of candidate structures at ambient conditions calculated using pure DFT and DFT+$U$ with LCM method are shown for (b) the U-Ga system and (d) the U-Al system. Experimental data for the U-Ga system are taken from Knudsen effusion mass spectrometry (KEMS) [29] and Calorimetry measurements [30], while experimental data for the U-Al system are taken from Ref. [31]. Dashed lines indicate unstable or metastable phases, whereas solid lines denote thermodynamically stable phases.

To further corroborate the validity of the LCM, we examine another system, the U-Al alloys, under ambient conditions. Ten representative stoichiometries are considered, yielding ten distinct candidate compounds. Pairwise combination of these compounds results in a total of $C_{10}^{2} = 45$ data tuples ($\Delta E$, $\Delta N_U$) as defined by Eq. (8). These data are plotted in Fig. 1(c). As expected, the data exhibit a clear linear relationship. The corresponding constant $\epsilon$ is determined to be $\epsilon$=1.1088, which is very close to the value obtained for the U-Ga system. In fact, the linearity between ($\Delta E$, $\Delta N_U$), as stated in Eq. (8), is not sensitive to the specific chemical system or pressure. Instead, it represents a general feature of the LCM, as further demonstrated for additional systems in Figs. S4–S6 of the SM.

The formation enthalpies and the corresponding convex hull obtained from the LCM are compared with those from pure DFT and available experimental data in Fig. 1(d) [31]. The comparison with plain DFT+$U$ is also illustrated in Fig. S2 of the SM. It is evident that neither plain DFT+$U$ nor pure DFT provides an adequate description of the U–Al system. Especially, plain DFT+$U$ fails to predict any stable $U_xAl_y$ compounds. Pure DFT fails to predict the stability of $UAl_3$ and $UAl_4$ at zero pressure, and falsely predicts that $U_2Al$ and UAl are stable. Even for $UAl_2$, the pure DFT predicted enthalpy is about 65.6% higher than the experimental data. In contrast, the LCM of DFT+*U* method correctly predicts the stability of $UAl_2$, $UAl_3$, and $UAl_4$ on the convex hull, which is in good agreement with the experimental observation. The predicted crystal structures of $MgNi_2$-type *P6₃/mmc*-$UAl_2$, *Pm-3m*-$UAl_3$ and *Imma*-$UAl_4$ are consistent with those reported in previous experimental studies [32-35]. Most importantly, the LCM-corrected formation enthalpies show excellent quantitative agreement with the experiments, with deviations of only 5.3% for $UAl_2$, and 4.1% and 6.2% for $UAl_3$ and $UAl_4$, respectively.

**B. Discovery of new high-pressure phases of U-M (M= Ga, Al, In) alloys by LCM**

As mentioned above, the LCM applied to DFT+*U* not only reproduces the correct phase stability and phase diagrams of the U-Ga and U-Al systems at ambient conditions, but also yields quantitatively accurate formation enthalpies. To the best of our knowledge, this is the first time that their phase stability and formation enthalpy are correctly predicted from pure first-principles, without any experimental input. Using this method, we further identified that the previously proposed *Cmcm*-$U_3Ga_5$ and $MgCu_2$-type *Pd-3m*-$UAl_2$ phases reported in earlier studies [33, 36, 37] are in fact metastable and not thermodynamically stable at zero pressure. As pointed out in Sec. II, LCM does not rely on pressure or other external conditions. Its demonstrative accuracy at zero pressure, together with the crucial advantage that no experimental data are required as input, enables reliable investigation of strongly correlated materials under extreme conditions, such as high pressure, where experimental energetic data are scarce or entirely unavailable. In this subsection, we focus on the important nuclear materials of U-M (M=Ga, Al, In) alloys. By combining first-principles structure predictions together with the LCM-corrected DFT+$U$ approach, we explore the most stable structures of $U_xGa_y$ compounds at pressures of 25, 50, 100, 150 and 200 GPa, respectively. The linearity of ($\Delta E$, $\Delta N_U$) tuple for all of these candidate structures of $U_xGa_y$ are checked for each pressure separately; and the constant $\epsilon$ for each pressure is obtained by least square fitting to Eq. (8) (Fig. S4).

The convex hull diagram of U-Ga system at 0 K, constructed using the formation enthalpies calculated with the LCM-corrected DFT+$U$ method, is shown in Fig. 2(a). Note that, here the proper magnetic ordering (FM, AFM, or NM) has been taken into account in all DFT+$U$ calculations, which is different from Fig. 1 where the temperature is room temperature and the magnetic state is NM. At zero pressure, the difference in magnetic ordering leads to a little deeper formation enthalpy, but does not affect the stability of competing structures. Figure 2(b) demonstrates the 0 K phase diagram of U-Ga system, in which the pressure ranges of the stable phases are shown. As can be seen, only the $AlB_2$-type *P6/mmm*-$UGa_2$ and $Cu_3Au$-type *Pm-3m*-$UGa_3$ phases are stable at zero pressure, which is in line with previous experimental studies [38, 39]. Upon increasing pressure, *P6/mmm*-$UGa_2$ phase becomes unstable at a pressure of 13 GPa, whereas the *Pm-3m*-$UGa_3$ phase transforms into a hexagonal *P6/mmm* phase at 44.7 GPa, which remains stable up to 82 GPa. For other Ga-rich compositions, we predict a previously unknown $U_2Ga_3$ compound that is stable between 45–200 GPa. It adopts a hexagonal structure with space group *P-62m*. On the U-rich side, a new $U_2Ga$ compound with the $AlB_2$-type *P6/mmm* structure is found to be stable between 108 GPa and 200 GPa.

The crystal structures of these newly predicted compounds are shown in Fig. 3. It is worth mentioning that the structure of *P6/mmm*-$U_2Ga$ compound adopts the same structure as those reported in U-based binary alloys with transition metals [40, 41]. The dynamical stability of all predicted stable intermetallic compounds is confirmed by their phonon spectra, which exhibit no imaginary frequencies (Fig. S8 in the SM).

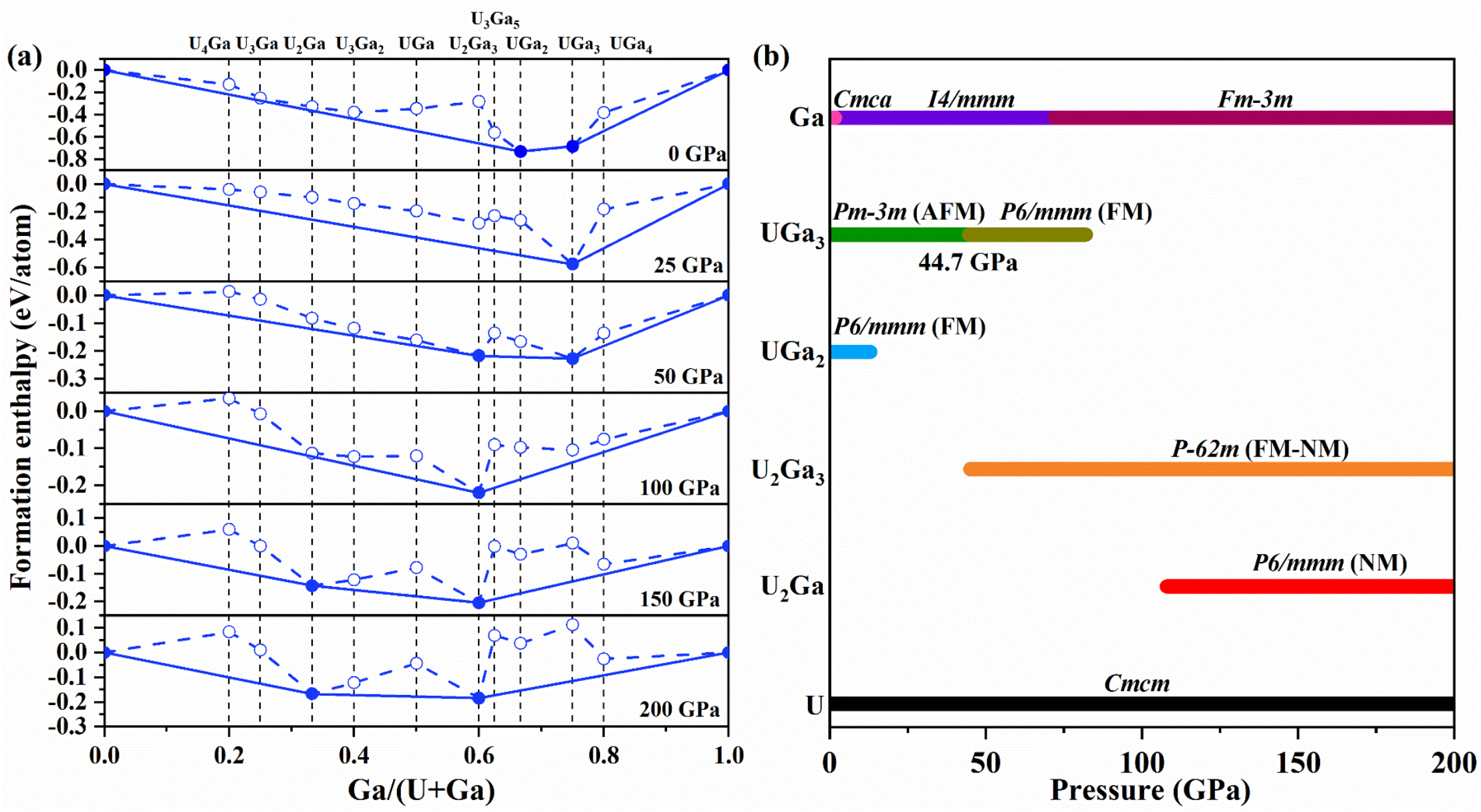

**Fig. 2** (color online) (a) Thermodynamic convex hull diagram of candidate structures in U-Ga system calculated by LCM of DFT+*U* at 0 K. Dashed-line connections indicate unstable or metastable phases, while solid-line connections indicate stable phases. (b) Pressure-composition phase diagram of U-Ga alloy in the range of 0-200 GPa and at 0 K predicted by LCM of DFT+*U*. Different color indicates different stable phases.

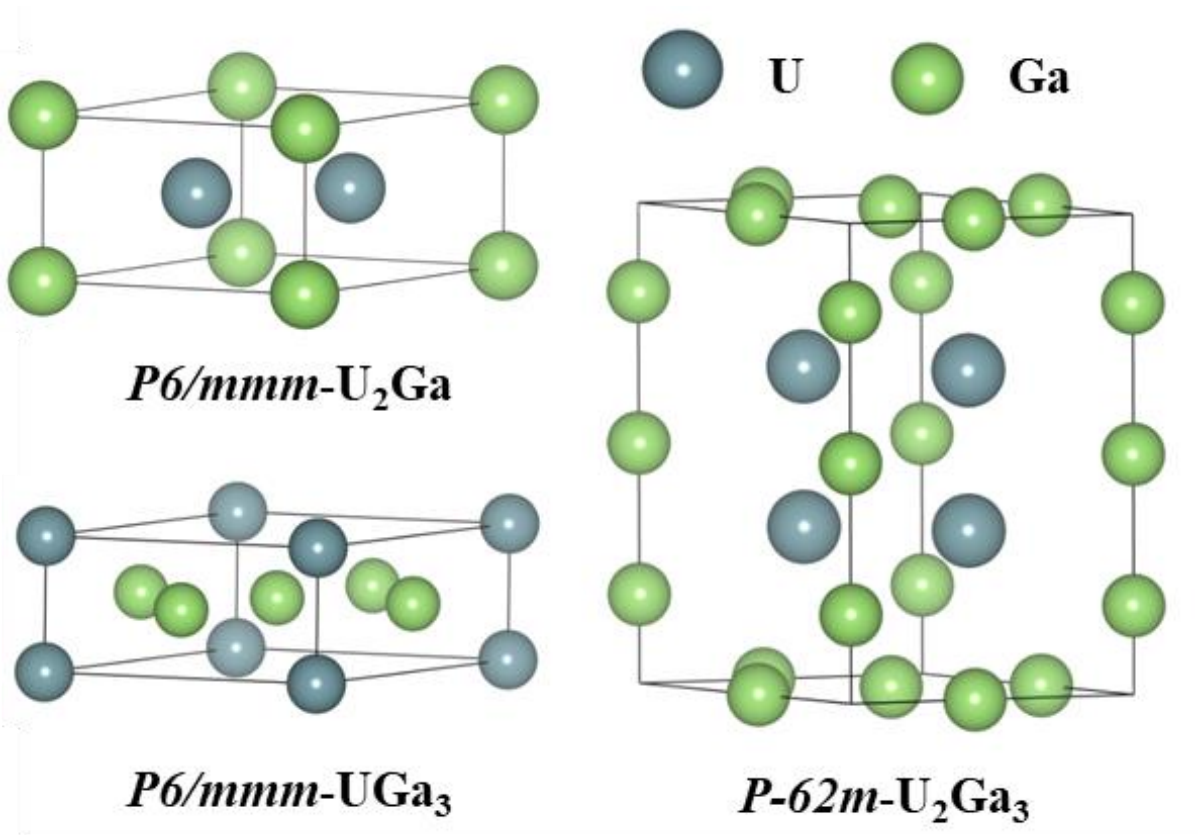

**Fig. 3** (color online) Structure of the newly predicted stable intermetallic compounds in U-Ga system.

Furthermore, the lattice parameters and magnetic moments of the $UGa_2$ and $UGa_3$ compounds calculated using DFT+*U* at 0 GPa and 0 K are compared with available low-temperature experimental data. Good agreement is obtained between theory and experiment. By contrast, pure DFT significantly underestimates the magnetic moments, as shown in Table 1. In addition, our study reveals that the previously proposed but controversial $Pu_3Pd_5$-type *Cmcm*-$U_3Ga_5$ phase is in fact not thermodynamically stable. It is quite close to the convex hull at zero pressure (with a distance of 0.12 eV/atom), and should be regarded as metastable. The calculated lattice parameters and magnetic moment of this $U_3Ga_5$ compound are also consistent with the reported experimental data (Table 1).

**Table 1.** Calculated lattice parameters and orbital ($\mu_L$) and spin ($\mu_S$) magnetic moment per uranium atom for *P6/mmm*-$UGa_2$, *Pm-3m*-$UGa_3$ and *Cmcm*-$U_3Ga_5$ phases at zero pressure, along with previous theoretical and experimental data, respectively.

| Phase | Method | Lattice parameters | Axis | Magnetic order | $u_L$ ($\mu_B$) | $u_S$ ($\mu_B$) | $\|u_L+u_S\|$ ($\mu_B$) |
|---|---|---|---|---|---|---|---|
| $UGa_2$ | PBE | a=b=4.230<br>c=3.990 | 100 | FM | -2.953 | 2.23 | 0.72 |
| | PBE+*U* | a=b=4.248<br>c=4.075 | 100 | FM | -3.662 | 1.553 | 2.11 |
| | Expt. [42] | a=b=4.213<br>c=4.020 | 100 | FM | — | — | 2.70 |
| $UGa_3$ | PBE | a=b=c=4.265 | 111 | AFM | -1.797 | 2.203 | 0.41 |
| | PBE+*U* | a=b=c=4.316 | 111 | AFM | -2.914 | 2.274 | 0.64 |
| | LDA+*U* [43] | — | 111 | AFM | -3.02 | 2.38 | 0.64 |
| | Expt. [44] | a=b=c=4.240 | 111 | AFM | — | — | 0.74 |
| $U_3Ga_5$ | PBE+*U* | a=9.502<br>b=7.789<br>c= 9.505 | 001 | FM | -0.953 | 2.235 | 1.28 |
| | Expt. [37, 38] | a=9.442<br>b=7.629<br>c=9.449 | 001 | FM | — | — | 1.1 |

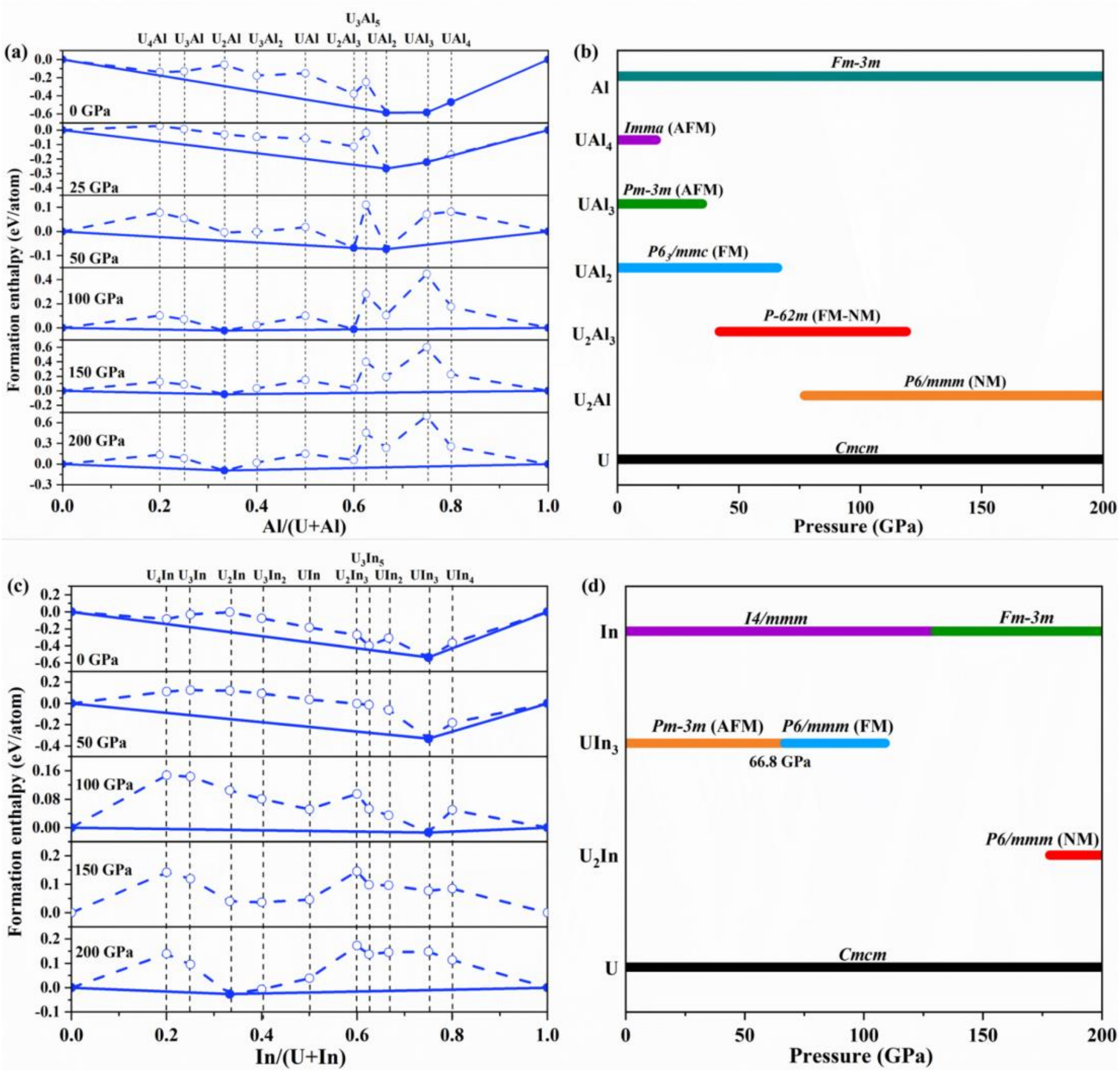

**Fig. 4** (color online) Thermodynamic convex hull diagrams of candidate structures for (a) U-Al and (c) U-In systems at different pressures and 0 K calculated with LCM for DFT+*U* method. The corresponding composition-pressure phase diagrams at 0 K for (b) U-Al and (d) U-In systems, in the range of 0–200 GPa, respectively.

We further apply the LCM to U-Al and U-In systems at high pressures up to 200 GPa. The calculated thermodynamic convex hull and composition-pressure phase diagrams of U-Al and U-In alloys at 0 K are shown in Fig. 4. At low temperature, all of these structures exhibit nontrivial magnetic orderings. At zero pressure, we find that $UAl_2$, $UAl_3$, and $UAl_4$ still remain thermodynamically stable at 0 K, but their magnetic orderings change from NM at room temperature to FM or AFM at 0 K. With increasing pressure, the *P6₃/mmc*-$UAl_2$, *Pm-3m*-$UAl_3$, and *Imma*-$UAl_4$ phases lose stability at 66, 35, and 16 GPa, respectively. In contrast, new compounds with different stoichiometries, namely *P6/mmm*-$U_2Al$ and *P-62m*-$U_2Al_3$, become stable in the pressure ranges of 77−200 GPa and 42−119 GPa, respectively.

Unlike the U-Al and U-Ga alloys, only one thermodynamically stable phase, *Pm-3m*-$UIn_3$, is found for the U–In system at zero pressure (Fig. 4(c, d)), in agreement with experimental observation [45]. With increasing pressure, this structure transforms into a hexagonal *P6/mmm* phase at 66.8 GPa, which remains stable up to 130 GPa. Notably, a similar structural transition is also predicted for $UGa_3$ (Fig. 2). At higher pressures, only one compound with the stoichiometry $U_2In$ and space group of *P6/mmm* is found to be stable, with a stability range of 178–200 GPa.

It is worth noting that although $UIn_3$ exhibits AFM ordering at low temperature, it becomes paramagnetic (PM) at room temperature. For this reason, we also examine the phase stability of the NM or PM states of $U_xIn_y$ compounds under ambient conditions. The corresponding ($\Delta E$, $\Delta N_U$) data tuple and the formation enthalpy calculated using the LCM of DFT+$U$ are obtained. Comparisons with the plain DFT+$U$, pure DFT and experimental data are presented in Fig. S3. Once again, the LCM-DFT+$U$ accurately reproduces the experimental formation enthalpy of the $UIn_3$ compound. In contrast, both pure DFT and plain DFT+$U$ incorrectly predict that no stable intermetallic compounds exist in the U-In system. This result further confirms the effectiveness and robustness of the LCM.

## IV. Discussion

In the preceding section, we applied the LCM to U-based binary alloys and demonstrated its effectiveness. In order to verify the broader applicability of LCM, we further explore other binary compounds beyond uranium alloys, as well as more complex ternary compound and surface adsorption problem. In the first case, we applied the LCM-DFT+*U* method to the Np-Al and U-Si systems, with the results shown in Fig. 5. For the Np-Al system, the formation enthalpies predicted by pure DFT deviate substantially from the reference values obtained from CALPHAD assessments [46], showing an average error of 67.8%; and the $NpAl_3$ compound is wrongly predicted as unstable. In contrast, the LCM method correctly predicts the phase stability of $NpAl_2$, $NpAl_3$, and $NpAl_4$, and significantly improves the theoretical prediction accuracy in formation enthalpy for these compounds, reducing the average error to only 9.7%. For the U-Si compounds, the DFT-predicted formation enthalpies of $U_3Si_2$, USi, and $USi_2$ are remarkably higher than experimental values and lie far above the convex hull, incorrectly indicating that these phases are unstable. The predicted average error for all compounds in this system is 28.5%. On the other hand, LCM substantially improves the results: it not only correctly predicted that $U_3Si_2$, USi, and $USi_2$ are stable, but also reduced the average error in the predicted formation enthalpies to only 9.2% compared to the experimental data.

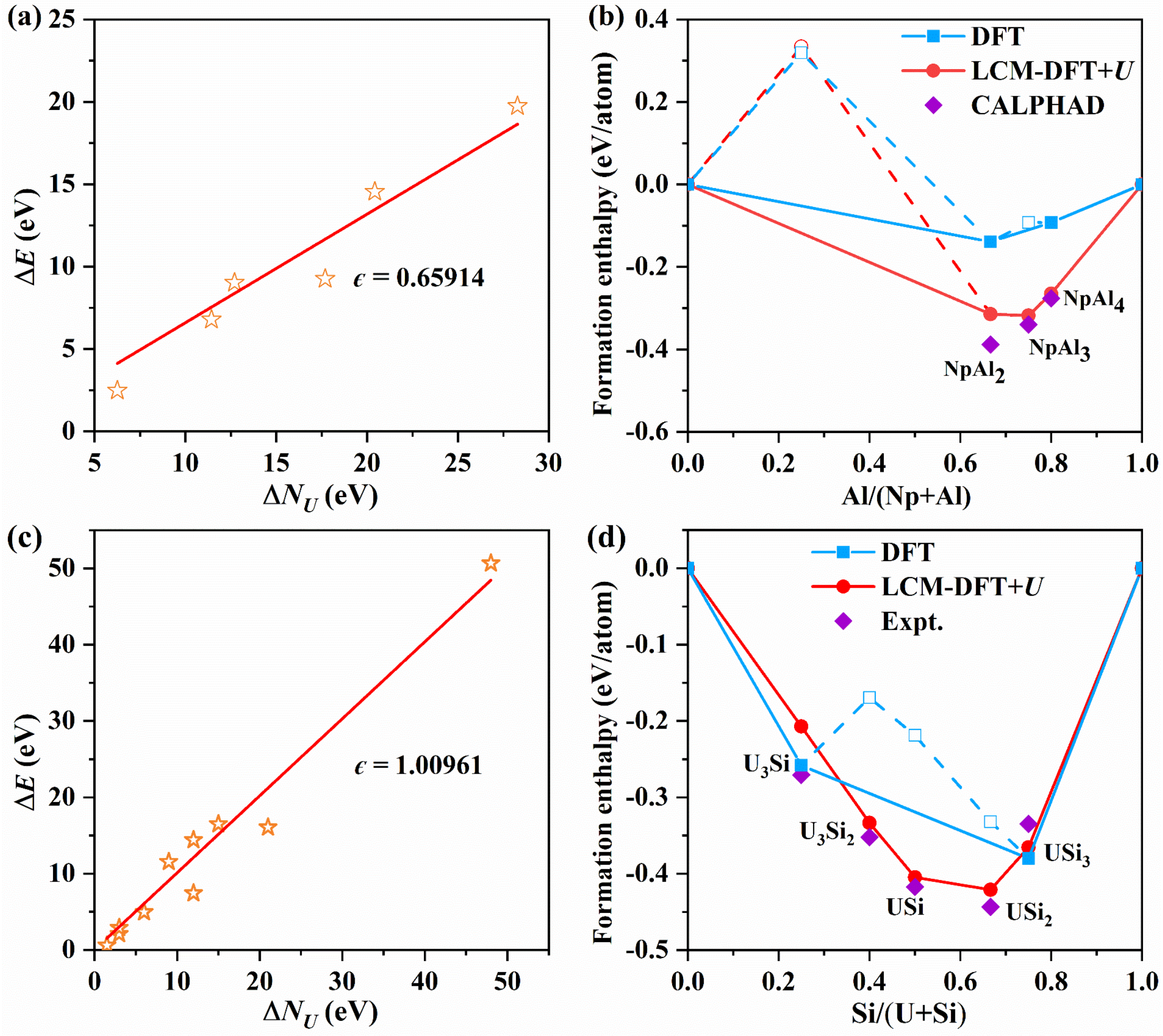

**Fig. 5** (color online) Calculated $\Delta E$-$\Delta N_U$ relationships and the correction constants for (a) Np-Al and (c) U-Si systems. Thermodynamic convex hull diagrams of candidate structures at ambient conditions calculated by DFT and LCM-DFT+$U$ methods: (b) Np-Al system, compared with the CALPHAD data [46]; (d) U-Si system, compared with the experimental data [47].

In the second case, we apply the LCM method to the non-actinide system, including bulk Cu-O system, oxygen adsorption on Cu surface, and a ternary system. For the Cu-O bulk system, the LCM-DFT+$U$ predicted formation enthalpies of CuO and $CuO_2$ compounds show great improvement over those obtained from pure DFT calculations (Table 2), with the LCM method yielding an average error of 8.4%—substantially lower than the 26.7% error of DFT. Beyond the bulk system, we also employ this method to investigate the adsorption energy of oxygen atom on the Cu(111) surface. Specifically, for an oxygen atom adsorbed at the fcc site on a clean Cu(111) (2×2 supercell) surface, the LCM-DFT+$U$ calculated adsorption formation enthalpy is

-4.901 eV. This value is slightly closer to the experimental value of -4.467 eV than the DFT result (-5.032 eV), corresponding to an error of 9.7% and 12.7%, respectively. Finally, we apply the LCM method to the ternary MnSnAu system, with the Hubbard term acts on the Mn atoms, for which the LCM also demonstrates excellent performance. The result is that the LCM-predicted formation enthalpy is -0.521 eV/atom, deviating by only 3.8% from the experimental value (-0.502 eV/atom). In contrast, the pure DFT gives a value of -0.053 eV/atom, showing a significantly larger error of 89.4% than LCM.

These additional case studies further demonstrate the power and the general applicability of LCM method, revealing that it can be extended beyond uranium-based binary alloys to include other strongly correlated systems, whether binary or ternary, even surface adsorption.

**Table 2.** Comparison of calculated formation enthalpies (in eV/O for Cu(111)-O, in eV/atom for other compounds) for different systems at zero pressure using pure DFT (GGA-PBE) and LCM-DFT+*U* methods, together with the corresponding reference data. The reference data for the Np-Al system are taken from the CALPHAD data [46], while those for the other compounds are experimental values in Refs. [24, 47-49].

| Compound | DFT | LCM-DFT+*U* | Reference data | Error (%) | |
|---|---|---|---|---|---|
| | | | | DFT | LCM |
| $NpAl_2$ | -0.139 | -0.315 | -0.388 | 64.2 | 18.8 |
| $NpAl_3$ | -0.092 | -0.318 | -0.340 | 72.9 | 6.5 |
| $NpAl_4$ | -0.093 | -0.265 | -0.277 | 66.4 | 4.3 |
| $U_3Si$ | -0.258 | -0.207 | -0.271 | 4.8 | 23.6 |
| $U_3Si_2$ | -0.170 | -0.333 | -0.352 | 51.7 | 5.4 |
| USi | -0.219 | -0.405 | -0.417 | 47.5 | 2.9 |
| $USi_2$ | -0.332 | -0.421 | -0.443 | 25.1 | 5.0 |
| $USi_3$ | -0.380 | -0.366 | -0.335 | 13.4 | 9.3 |
| CuO | -0.938 | -0.832 | -0.820 | 14.4 | 1.5 |
| $CuO_2$ | -0.806 | -0.669 | -0.580 | 39.0 | 15.3 |
| Cu(111)-O | -5.032 | -4.901 | -4.467 | 12.7 | 9.7 |
| MnSnAu | -0.053 | -0.521 | -0.502 | 89.4 | 3.8 |

## V. Conclusion

A general LCM that is fully *ab initio* to correct the energy (as well as the pressure and enthalpy) of DFT+*X* methods introduced by the Hubbard model was proposed, which effectively removed the unphysical dependence of the calculated energetics on the model parameters without requiring any experimental data as input. The validation and accuracy of this method were demonstrated using the framework of DFT+*U* for important nuclear materials of U-Ga, U-Al, and U-In systems under ambient conditions. Correct phase stability and lattice structures were obtained. The accuracy in the calculated formation enthalpy of stable phases has been significantly improved from the totally unphysical results of plain DFT+*U* to within 6% of the experimental values using the LCM. This perfect reproduction of the experimental formation enthalpy and phase stability of the strongly correlated alloys with a fully first-principles approach is very impressive, and is the first time to the best of our knowledge.

A key feature of the LCM is that it requires no experimental input, which greatly expands the scope for investigating the phase stability and phase diagrams of strongly correlated materials under extreme conditions. In such regimes, DFT+*X* methods and accurate first-principles formation enthalpies are essential, while experimental energetic data are often unavailable. To illustrate this capability, the LCM of DFT+*U* was applied to high-pressure phases of U-Ga, U-Al, and U-In systems up to 200 GPa. Three new phases were discovered at high pressures: (*i*) *P6/mmm* phases of $U_2Ga$, $U_2Al$, and $U_2In$; (*ii*) *P-62m* phases of $U_2Ga_3$ and $U_2Al_3$; and (*iii*) *P6/mmm* phases of $UGa_3$ and $UIn_3$, respectively. Our results indicate that under high pressure, not only the crystal structures but also the compositions of all known low-pressure intermetallic phases of U-Ga ($UGa_3$ and $UGa_2$), U-Al ($UAl_2$, $UAl_3$, $UAl_4$), and U-In ($UIn_3$), lose their stability and decompose. In contrast, new stoichiometries, including $U_2Ga_3$/$U_2Al_3$ and $U_2Ga$/$U_2Al$/$U_2In$, emerge and become stable. These insights unveiled a completely new physical picture about the uranium based binary alloys with main group elements, and demonstrated the power of the LCM. Furthermore, the broad applicability of LCM has been validated across a range of diverse systems, including binary systems such as Np-Al, U-Si, and Cu-O, as well as in more complex cases such as oxygen adsorption on the Cu(111) surface and the ternary MnSnAu compound. Since applying LCM with other DFT+*X* methods is similar to DFT+*U*, and is rather straightforward, progress along this line will open a new avenue for accurate description of strongly correlated systems, and will accelerate the discovery of novel correlated materials.

## Method and Computational details

Variable-composition structure searches for U-M (M=Ga, Al, In) systems were carried out at five different pressures of 0, 50, 100, 150 and 200 GPa to generate candidate structures, using the evolutionary algorithm as implemented in USPEX package [50]. The unit cell contains up to 24 atoms. The first generation includes 100 random structures and for the remaining generations, each contains 60 structures. A total of 60 generations were generated for every search. The best ten candidate structures for each composition with the lowest enthalpy were then re-optimized with higher computational accuracy to find the most stable structure. Structural relaxation and total energy calculations were performed using DFT [1, 2] as implemented in the Vienna Ab initio Simulation Package (VASP) [51, 52]. The exchange-correlation functional is described by the Perdew-Burke-Ernzerhof (PBE) [53] parameterization of the GGA. The atomic coordinates and lattice vectors were fully optimized until the Hellmann–Feynman forces acting on each atom were less than $10^{-3}$ eV/Å and the convergence of the total energy was better than $10^{-7}$ eV between two successive ionic steps. The cut-off energy of 550 eV for the plane-wave basis set and the Monkhorst-Pack [54] k-point meshes with a grid spacing of $2\pi \times 0.02$ Å$^{-1}$ were used to ensure that the energy of each structure is converged to 1 meV per atom or better for all calculations. In order to ensure the dynamic stability of the predicted compounds, the phonon dispersion curves were checked by using the small-displacement method as implemented in MyPhon [55].

For systems containing *d* or *f* electrons that exhibit strong electronic correlations, DFT+*U* approach was employed to deal with the on-site coulomb interactions between the *f* electrons of uranium atoms in U-M (M=Al, Ga, In) compounds. Specifically, we used the simplified DFT+*U* method proposed by Dudarev et al. [11] with only a single effective *U* parameter. Linear response method proposed by Cococcioni et al. [56] was used to determine the effective Hubbard *U* value for all U-M compounds through both self-consistent and non-self-consistent calculations within the framework of the PBE, thus renders our DFT+*U* as fully *ab initio*. The consequence is that the Hubbard *U* parameter might be different for different U-M compounds at different pressure, and LCM provides the theoretical basis to handle this variation. Spin polarization was taken into account in different magnetic ordering, including antiferromagnetic (AFM), ferromagnetic (FM), and non-magnetic (NM) ordering. For U-In system, we also considered paramagnetic (PM) ordering. The possible metastable electronic states that is notorious in DFT+*U* calculations were eliminated by using QA method [57]. It should be pointed out that in order to reduce the computational demanding for DFT+*U* calculations, the initial candidate structures were generated by using USPEX with PBE functional. These structures were then screened by structure relaxations with DFT+*U* and the formation enthalpies calculated by plain DFT+*U* or LCM for DFT+*U*, respectively. The obtained structure and energetics, by using pure DFT, plain DFT+*U*, and LCM for DFT+*U*, were then compared with each other.

The same first-principles calculation methods were applied to the systems of Np-Al, U-Si, Cu-O, MnSnAu, and oxygen adsorption on Cu(111) surface, except for them we mainly focused on LCM performance on the ambient structures, and did not perform structure search at high pressures.

**Acknowledgments**

This work was supported by National Key R&D Program of China under Grant No. 2021YFB3802300, the fund of national key laboratory of shock wave and detonation physics under Grant No. 2023JCJQLB05401, the NSAF under Grant Nos. U1730248 and U1830101, the National Natural Science Foundation of China under Grant Nos. 12372370, 12202418, 12074274, the Doctoral Research Fund of Southwest University of Science and Technology (Grant No. 25zx7147). Part of the simulation was performed on resources provided by the center for Comput. Mater. Sci. (CCMS) at Tohoku University, Japan.

**Author Contribution**

Xiao L. Pan: Calculation, Analysis, Writing, Original draft preparation. Hong X. Song: Analysis, Reviewing and Editing. Y. Sun: Analysis, Reviewing. F. C. Wu: Analysis, Writing. H. Wang: Analysis, Reviewing. Y. Chen: Analysis, Reviewing. Xiang R. Chen: Writing, Reviewing and Editing, Fund. Hua Y. Geng: Idea conceiving, Project design, Theory formulating, Writing, Reviewing and Editing, Fund.

TOC

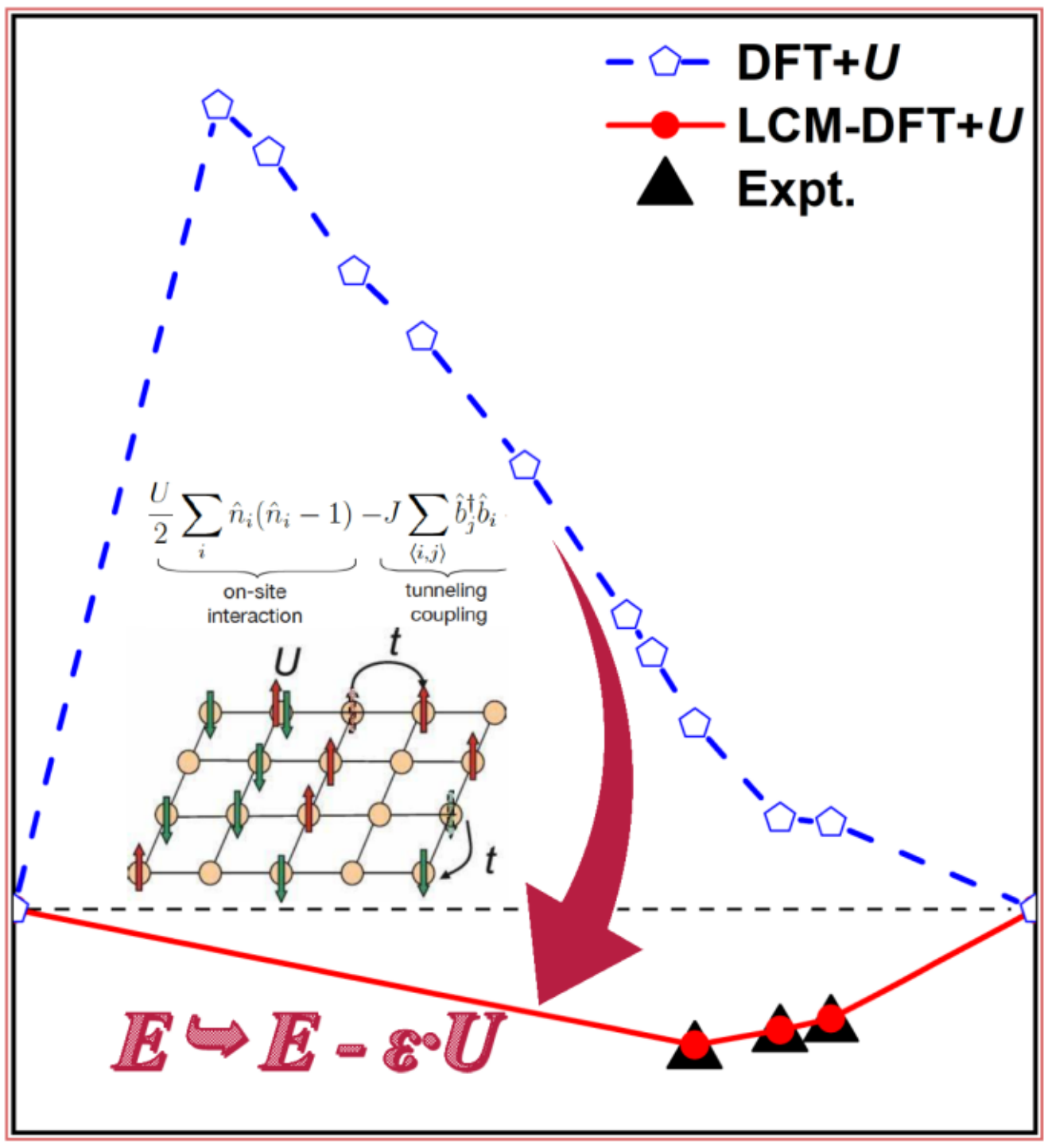

Hubbard model is pivotal to describe strongly correlated materials. But when incorporated with DFT to form DFT+$U$, or generally DFT+$X$ method, its energy becomes dependent on the model parameters and not well-defined. This notorious problem leads to unphysical formation energy. Linear correction method proposed here successfully eliminate the ambiguity in the energetics of DFT+$X$, and gives rise to the accurate formation enthalpy that is in excellent agreement with experimental data for the first time. This progress allows self-consistently varying of the model parameters, thus unprecedentedly makes DFT+$U$ a purely ab initio method.

# Supplementary Material for "General linear correction method for DFT+*X* energy: application to U-M (M=Al, Ga, In) alloys under high pressure"

Xiao L. Pan[1,2,3], Hong X. Song[1], Y. Sun[1], F. C. Wu[1]*, H. Wang[1], Y. F. Wang[1], Y. Chen[5], Xiang R. Chen[2]*, Hua Y. Geng[1,4]*

[1] *National Key Laboratory of Shock Wave and Detonation Physics, Institute of Fluid Physics, China Academy of Engineering Physics, Mianyang, Sichuan 621900, P. R. China;*

[2] *College of Physics, Sichuan University, Chengdu 610065, P. R. China;*

[3] *Extreme Condition Material Properties Science and Technology center, School of Mathematics and Physics, Southwest University of Science and Technology, Mianyang, 621010, P. R. China;*

[4] *HEDPS, Center for Applied Physics and Technology, and College of Engineering, Peking University, Beijing 100871, P. R. China;*

[5] *Fracture and Reliability Research Institute, School of Engineering, Tohoku University, Sendai 980-8579, Japan.*

## SI. Supplementary computational details

In order to determine the dynamic stability of the predicted intermetallic compounds, the phonon dispersion curves were calculated using the small-displacement method implemented in MyPhon code [1]. A 2 × 2 × 4 supercell for *P6/mmm*-$U_2M$ (48 atoms), 2 × 1 × 2 supercell for $U_3M$ (64 atoms), 2 × 2 × 1 supercell for *Immm*-$U_4M$ (40 atoms), 2 × 2 × 1 supercell for *Pmma*-$U_3M_2$ (40 atoms), 2 × 2 × 1 supercell for *P-62m*-$U_2M_3$ (40 atoms), 2 × 2 × 2 supercell for $UM_3$ (32 atoms), 2 × 2 × 1 supercell for *P6₃/mmc*-$UAl_3$ (96 atoms), and 2 × 2 × 1 supercell for *Imma*-$UAl_4$ (80 atoms) were employed in the calculations of phonon spectra.

The linear response method proposed by Cococcioni *et al.* [2] was used to theoretically and self-consistently determine the Hubbard *U* value for all U-M compounds at each pressure through both self-consistent and non-self-consistent calculations within the framework of PBE in VASP. This renders our DFT+$U$ as fully *ab initio* (in the sense that does not require any experimental or empirical input), and the specific $U$ value varies with structure, composition and pressure. In the linear response method, by applying an additional spherical potential to a single uranium atom in U-M compounds, the $f$-orbital charge density of the uranium atom is redistributed. Thus, both self-consistent and non-self-consistent response functions can be obtained by following equations:

$$\chi = \left(\frac{\partial N^{\mathrm{scf}}}{\partial V}\right), \quad \chi_0 = \left(\frac{\partial N^{\mathrm{nscf}}}{\partial V}\right) \tag{1}$$

where $V$ represents the applied additional spherical potential, $N^{\mathrm{scf}}$ and $N^{\mathrm{nscf}}$ represent the $f$-electrons number of uranium atom obtained via self-consistent and non-self-consistent calculations under the additional spherical potential, respectively. With the non-self-consistent and the self-consistent response functions, the self-consistent value of $U$ parameter for the DFT+$U$ treatment of the uranium $f$-electrons in U-M compounds is obtained by the following equation:

$$U = \chi^{-1} - \chi_0^{-1} = \left(\frac{\partial N^{\mathrm{scf}}}{\partial V}\right)^{-1} - \left(\frac{\partial N^{\mathrm{nscf}}}{\partial V}\right)^{-1} \tag{2}$$

* *Corresponding authors. E-mail: s102genghy@caep.cn; xrchen@scu.edu.cn; fcwu2011@mail.ustc.edu.cn*

In order to get accurate result, six additional potentials (-0.1, -0.05, -0.025, 0.025, 0.05, and 0.1 eV) were used for the self-consistent and non-self-consistent calculations. The corresponding response function was then obtained by linearly fitting the $f$-electrons number of uranium atom as a function of the additional potential $V$.

In the DFT+$U$ calculations for the Np – Al system, the Hubbard $U$ parameters for each compound were determined using the linear response method. The resulting $U$ values are 2.618 eV for $Np_3Al$, 2.997 eV for $NpAl_2$, 3.140 eV for $NpAl_3$, and 3.135 eV for $NpAl_4$. For the U – Si, Cu – O, and MnSnAu systems, a single Hubbard $U$ value was employed for all compounds within each system for brevity, namely 1.5 eV for U – Si, 4.0 eV for Cu – O, and 4.0 eV for MnSnAu. In our calculations, the $O_2$ energy is corrected following Wang et al. [3], which removes oxygen-related GGA errors and enables a reliable assessment of the Hubbard $U$ correction applied to the Cu $d$ orbitals.

# SII. Supplementary figures

## A. The Hubbard *U* parameters used in DFT+*U* calculations

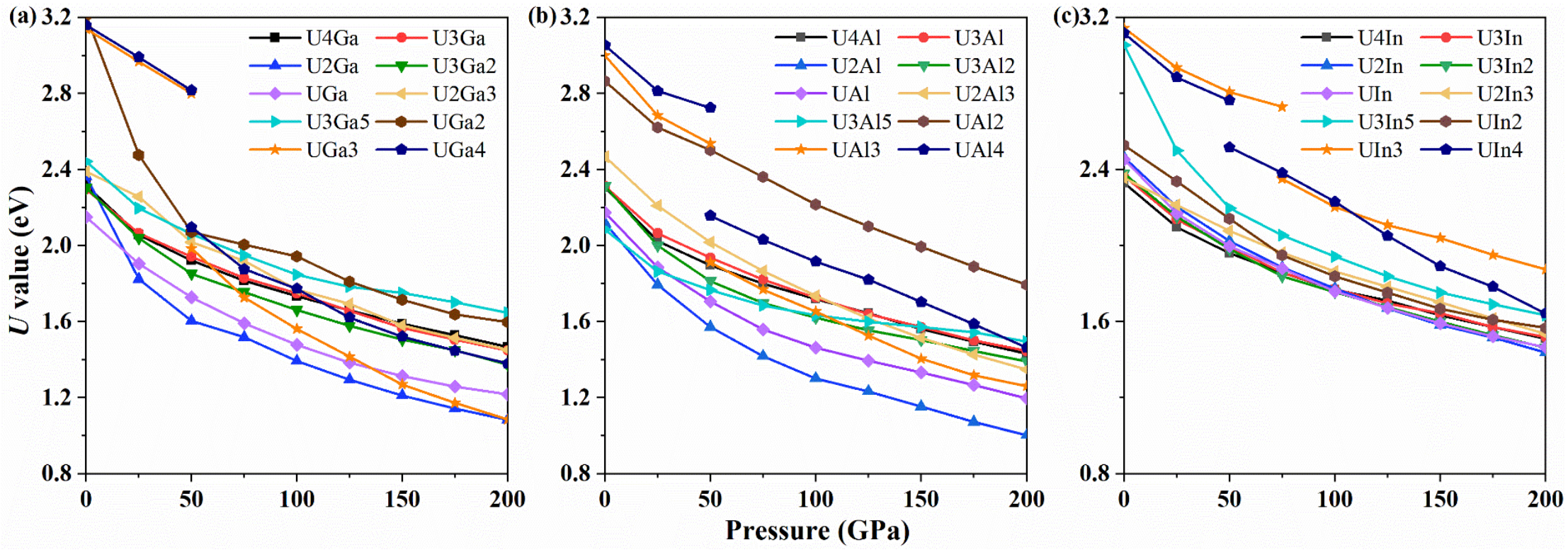

**Fig. S1** (color online) Variation of the calculated Hubbard *U* parameter as a function of pressure for all U-M compounds. They are used in subsequent DFT+*U* calculations. (a) U-Ga system; (b) U-Al system; (c) U-In system. Note that these *U* parameters are obtained via the linear response method. For $UM_3$ and $UM_4$ (M=Ga, Al, In) compositions, the jump in *U* indicates that the structure is changed at that pressure.

## B. Comparison of formation enthalpies calculated by LCM-DFT+*U* and plain DFT+*U* methods

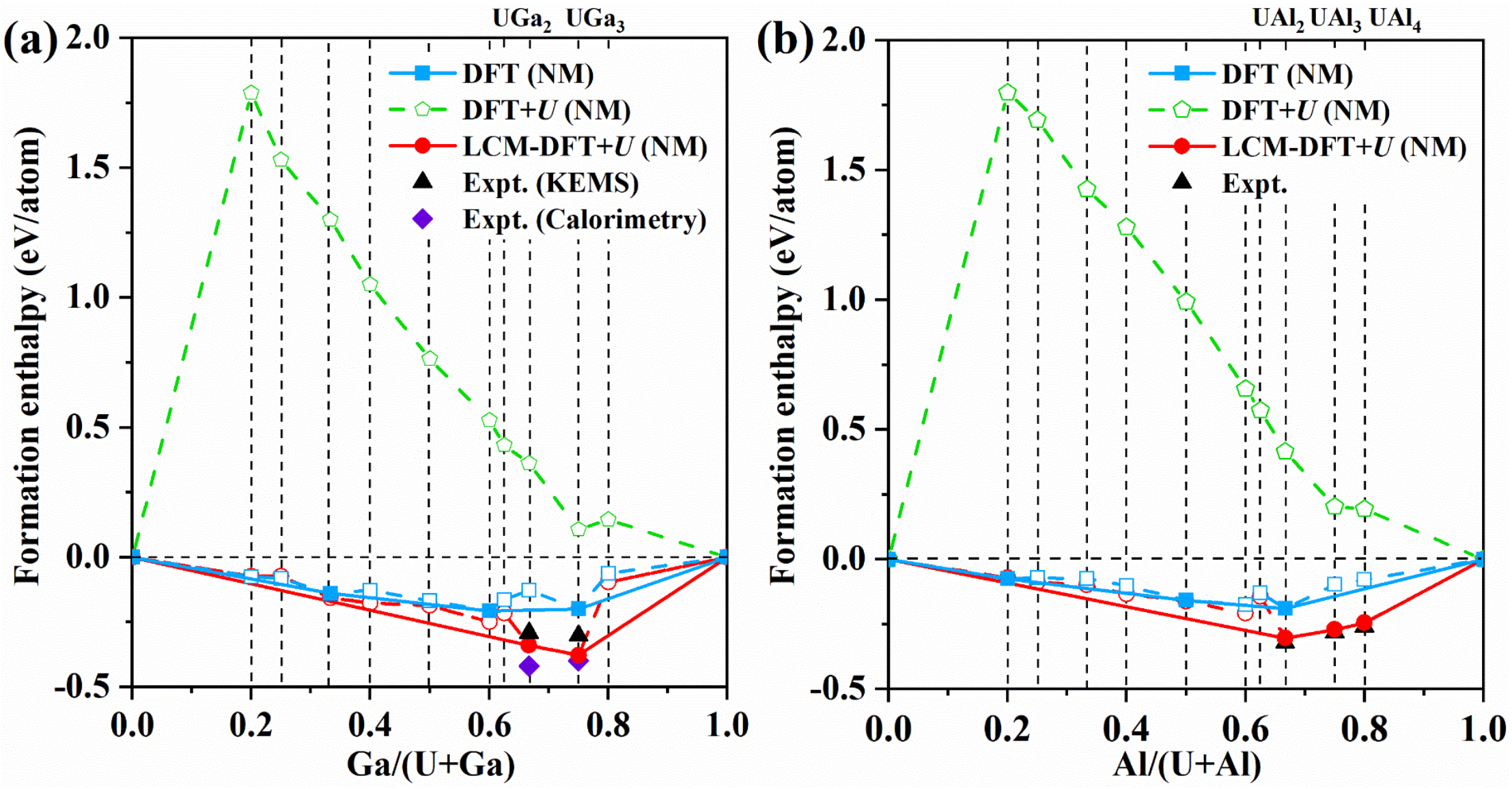

**Fig. S2** (color online) Comparison of the thermodynamic convex hull diagrams of candidate structures at ambient conditions calculated by pure DFT, plain DFT+*U*, and LCM-DFT+*U* methods: (a) U-Ga system, along with experimental data taken from Knudsen Effusion Mass Spectrometry (KEMS) [4] and Calorimetry [5] for comparison; (b) U-Al system, along with the experimental data [6]. Dashed-line connections indicate unstable or metastable phases, while solid-line connections indicate stable phases. Here the "plain DFT+$U$" means the formation enthalpy is calculated directly using Eq. (5) of the main text with the energetics of DFT+$U$ without any correction.

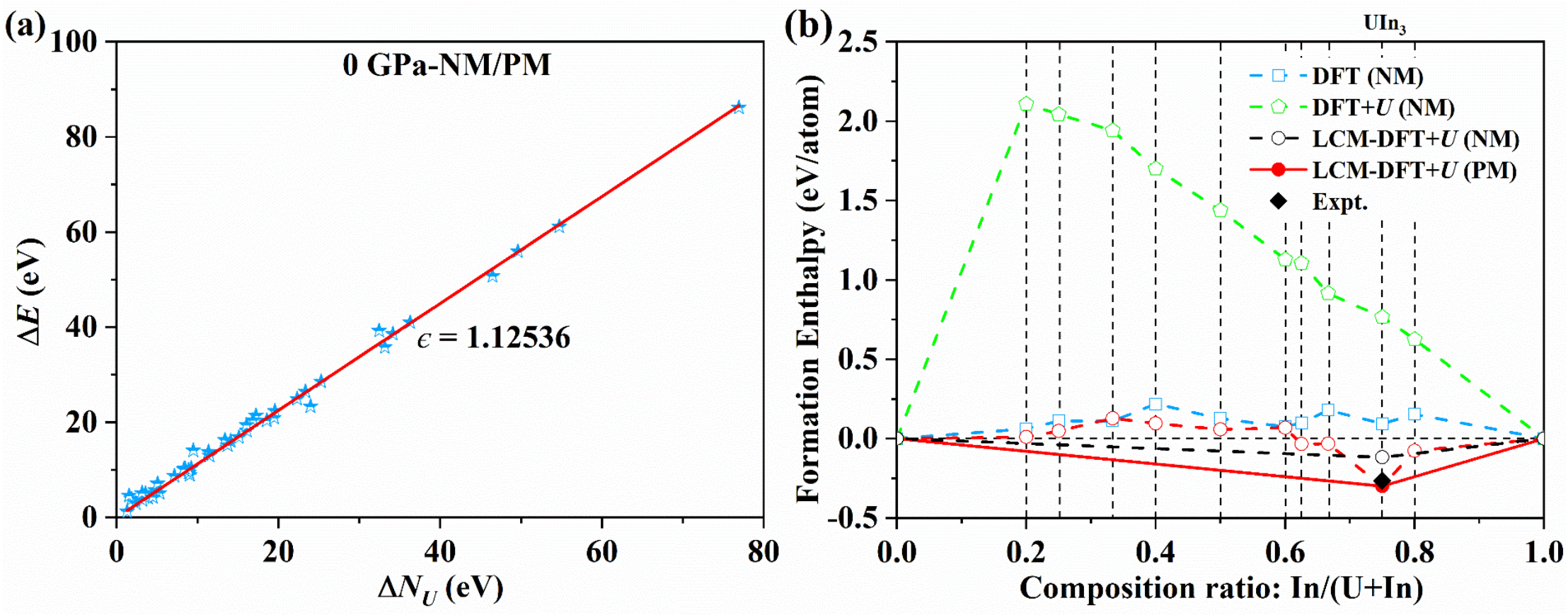

**Fig. S3** (color online) (a) The $\Delta E$-$\Delta N_U$ relationship for U-In system at ambient conditions, considering paramagnetic ordering for *Pm-3m*-$UIn_3$ and nonmagnetic ordering for other compositions. For the paramagnetic phase calculation of *Pm-3m*-$UIn_3$, we used a 2×2×2 supercell (32 atoms), fixing the magnetic moment of each atom at 2.75 $\mu_B$ but with random directions, while ensuring the total magnetic moment of the system remains zero. Eight sets of random magnetic moment directions were considered, and the averaged energy of these configurations was used to calculate the formation enthalpy of the *Pm-3m*-$UIn_3$ in paramagnetic state. (b) Comparison of the thermodynamic convex hull diagrams of candidate structures at ambient conditions calculated by pure DFT, plain DFT+$U$, and LCM-DFT+$U$ methods with the experimental data [6], respectively. Here the NM and PM denote nonmagnetic and paramagnetic ordering, respectively.

### C. The $\Delta E$-$\Delta N_U$ relationship of U-M compounds at different pressures

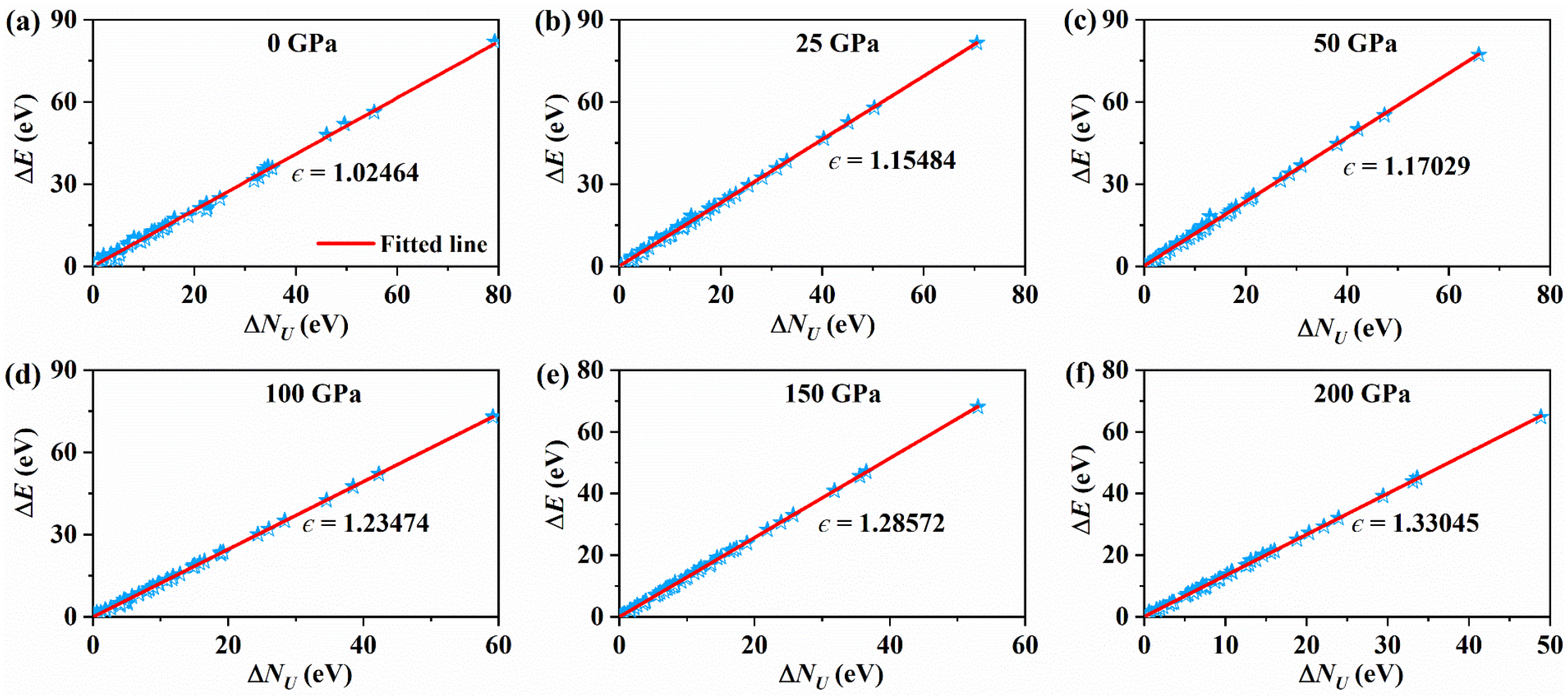

**Fig. S4** The $\Delta E$-$\Delta N_U$ relationship for U-Ga system at 0 K and at pressures of 0, 25, 50, 100, 150, and 200 GPa, respectively. The magnetic ordering (NM, FM or AFM) has been taken into account.

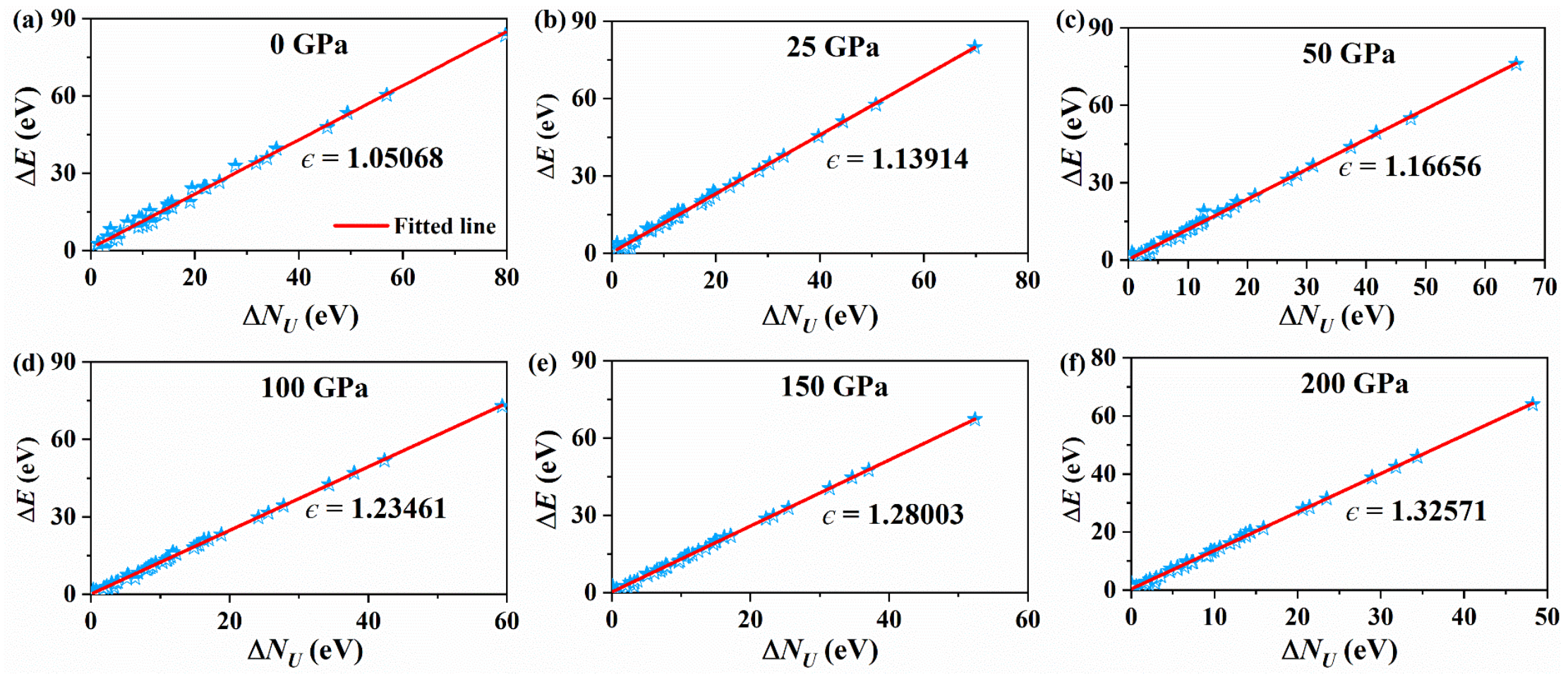

**Fig. S5** The $\Delta E$-$\Delta N_U$ relationship for U-Al system at 0 K and at pressures of 0, 25, 50, 100, 150, and 200 GPa, respectively. The magnetic ordering (NM, FM or AFM) has been taken into account.

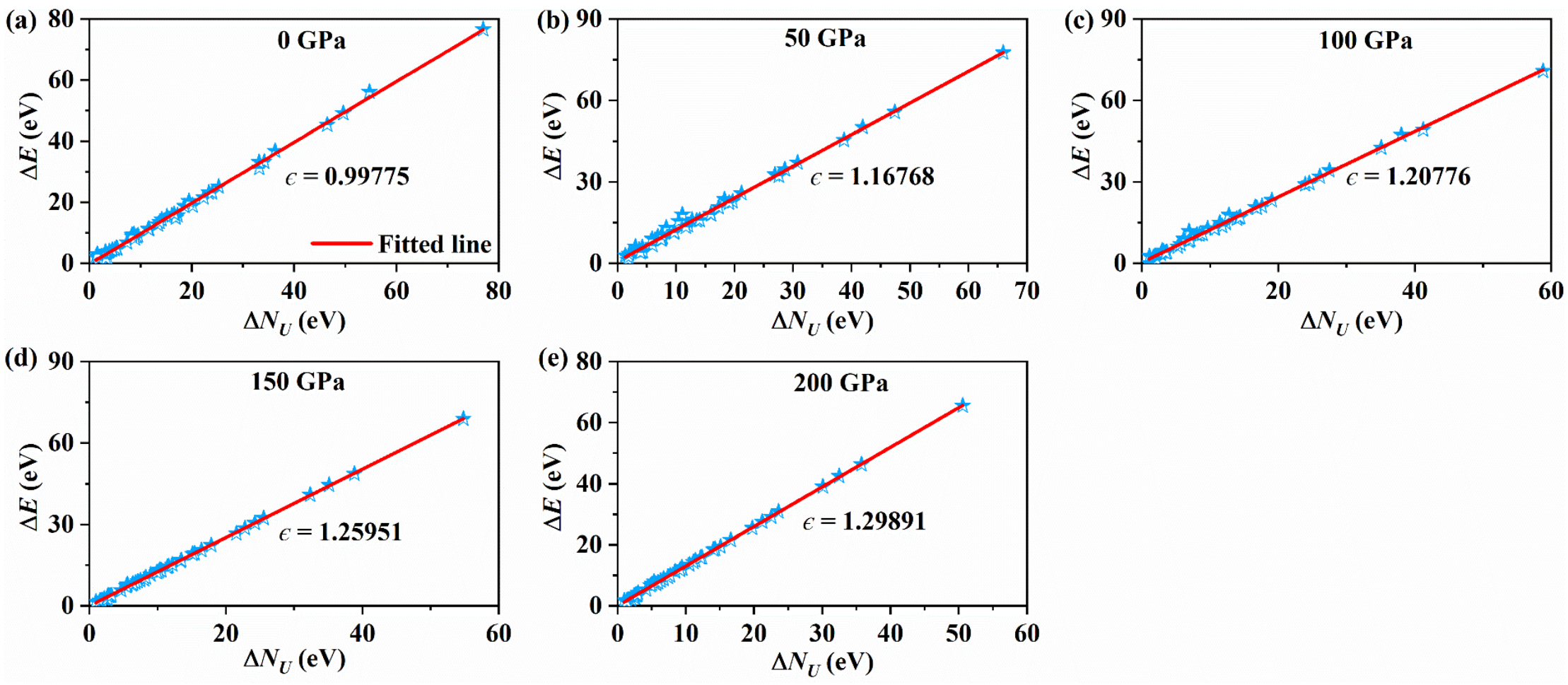

**Fig. S6** The $\Delta E$-$\Delta N_U$ relationship for U-In system at 0 K and at pressures of 0, 50, 100, 150, and 200 GPa, respectively. The magnetic ordering (NM, FM or AFM) has been taken into account.

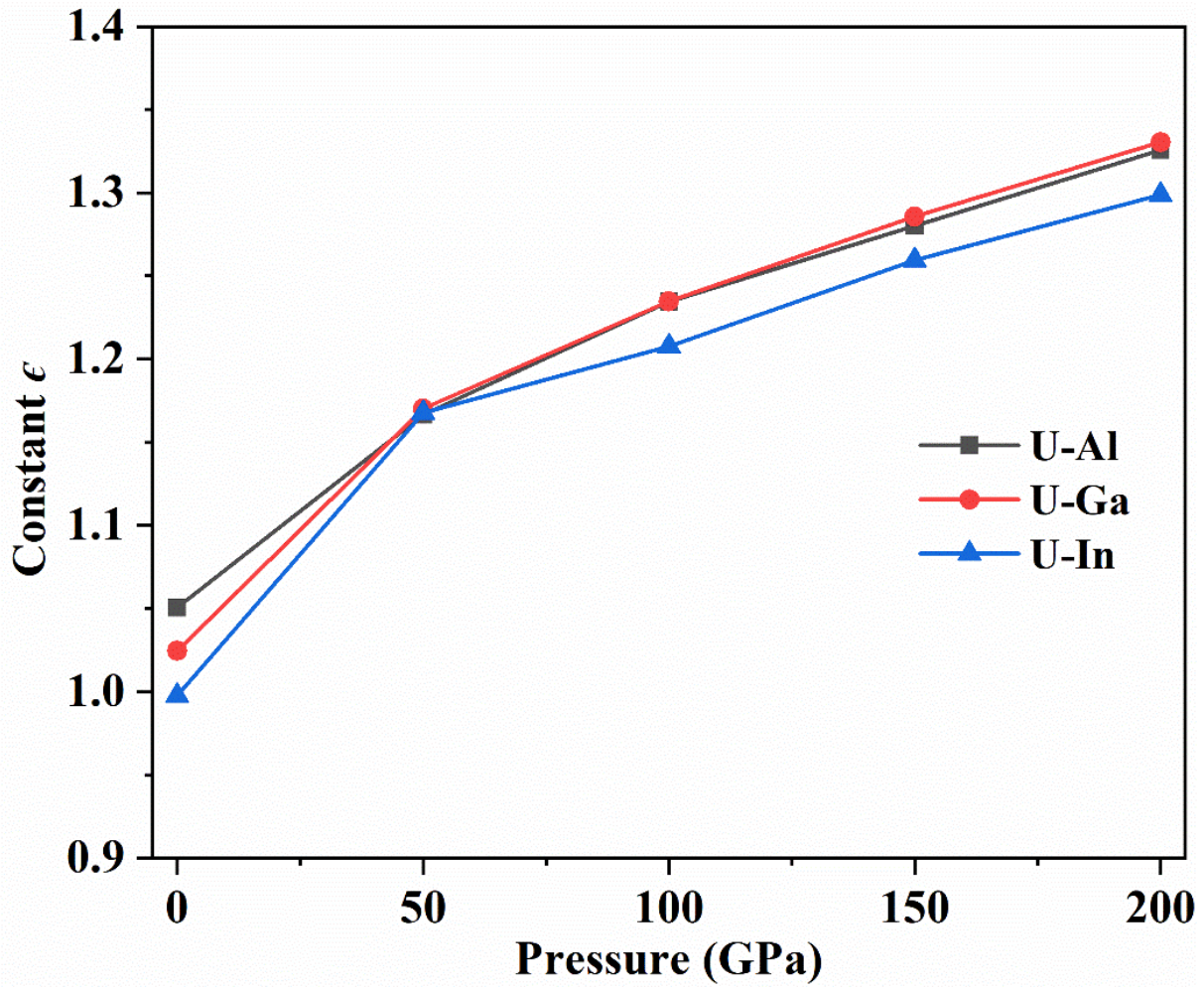

**Fig. S7** Variation of the constant $\epsilon$ with pressure for U-Al, U-Ga, and U-In systems, as listed in Figs. S4-S6, respectively.

## D. Phonon spectra

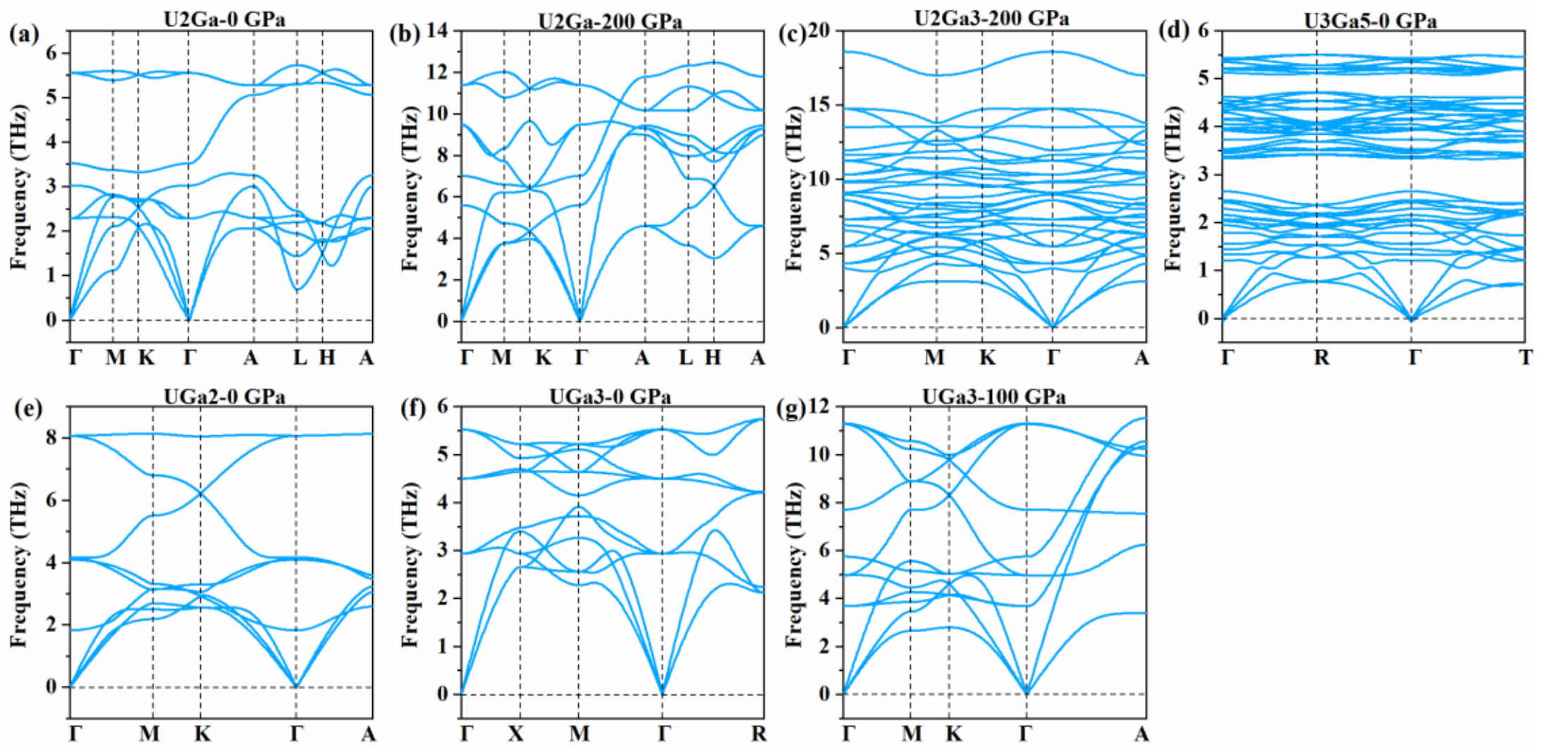

**Fig. S8** Calculated phonon spectra for *P6/mmm*-$U_2Ga$ at (a) 0 GPa and (b) 200 GPa, (c) *P-62m*-$U_2Ga_3$ at 200 GPa, (d) *Cmcm*-$U_3Ga_5$ at 0 GPa, (e) *P6/mmm*-$UGa_2$ at 0 GPa, (f) *Pm-3m*-$UGa_3$ at 0 GPa, and (g) *P6/mmm*-$UGa_3$ at 100 GPa, respectively.

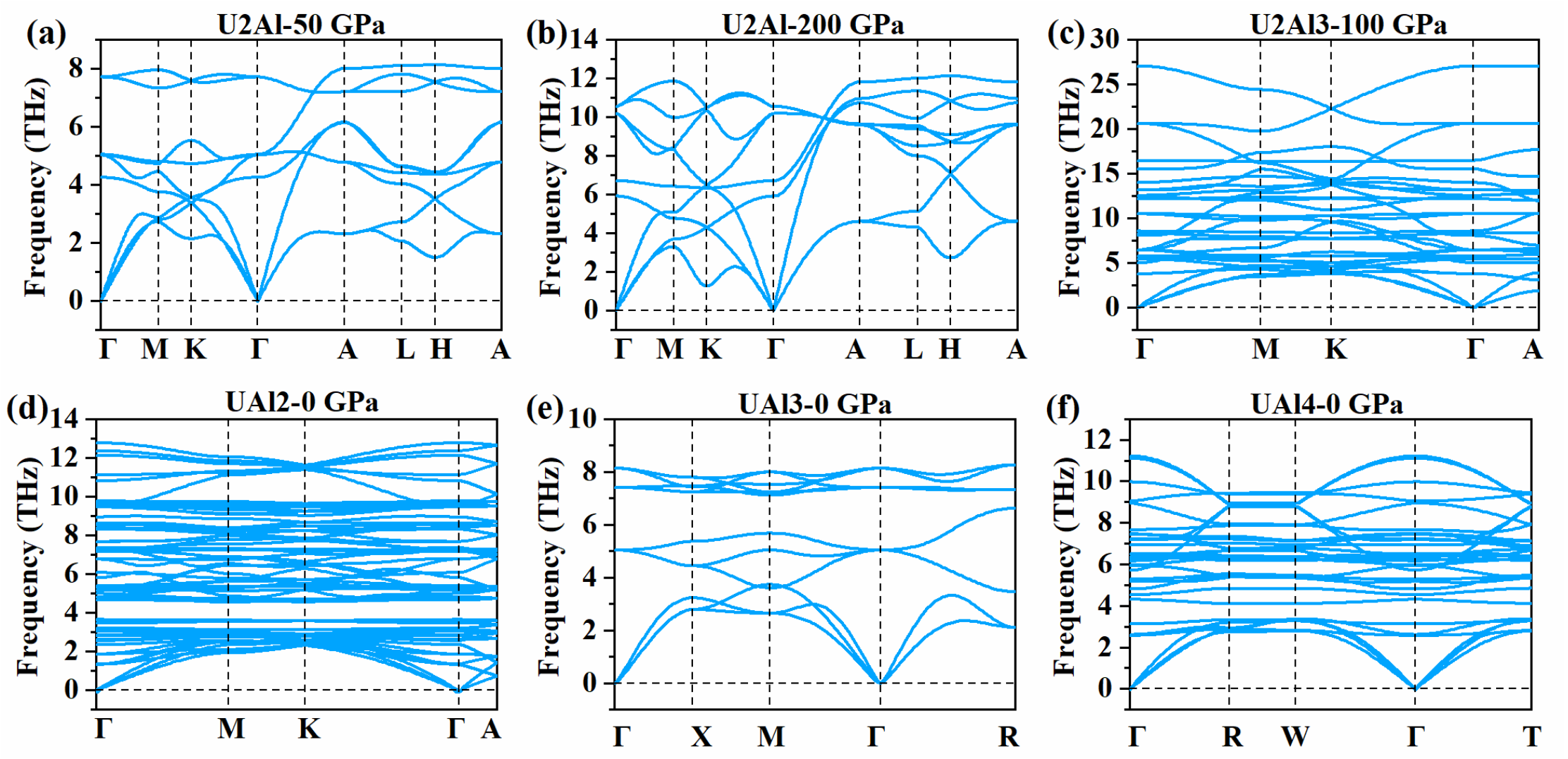

**Fig. S9** Calculated phonon spectra for *P6/mmm*-$U_2Al$ at (a) 50 GPa and (b) 200 GPa, (c) *P-62m*-$U_2Al_3$ at 100 GPa, (d) *P6₃/mmc*-$UAl_2$ at 0 GPa, (e) *Pm-3m*-$UAl_3$ at 0 GPa, and (f) *Imma*-$UAl_4$ at 100 GPa, respectively.

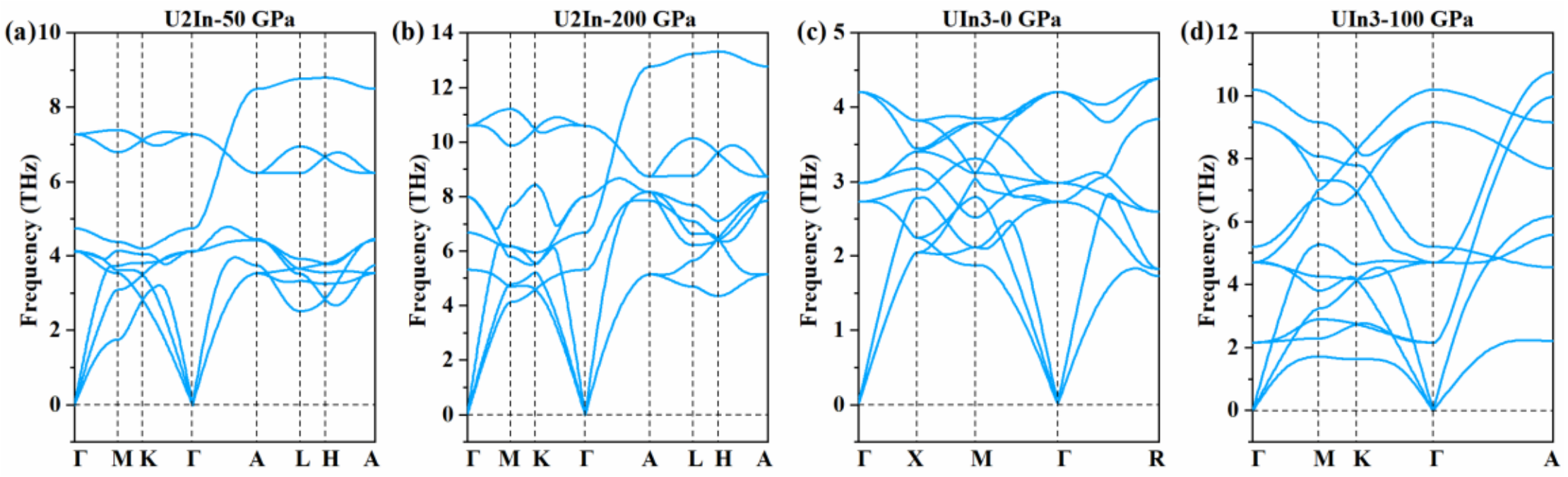

**Fig. S10** Calculated phonon spectra for *P6/mmm*-$U_2In$ at (a) 50 GPa and (b) 200 GPa, (c) *Pm-3m*-$UIn_3$ at 0 GPa, and (d) *P6/mmm*-$UIn_3$ at 100 GPa, respectively.

## E. Electronic structure and bonding properties

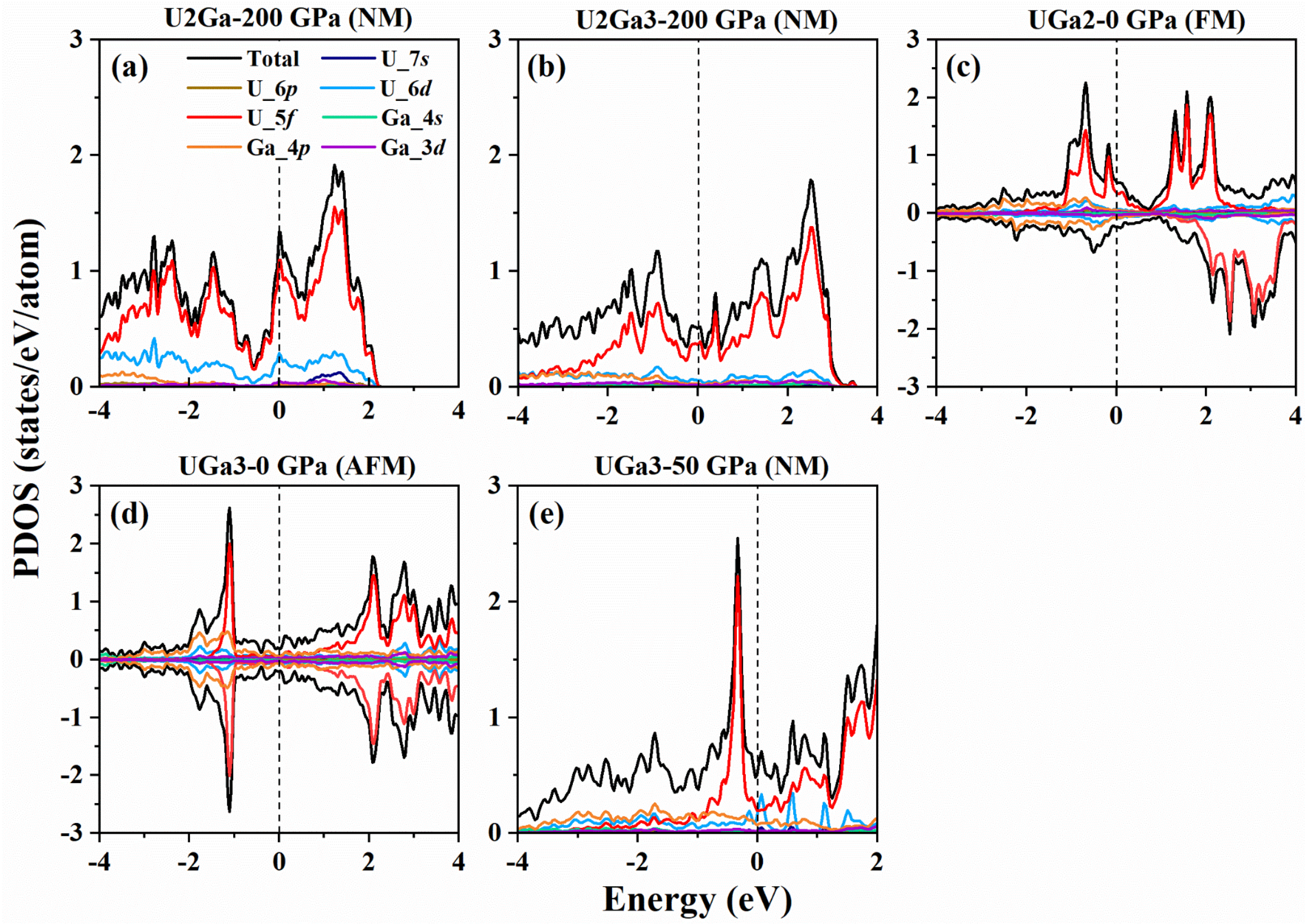

**Fig. S11** (color online) Calculated electronic projected density of states (PDOS) for (a) *P6/mmm*-$U_2Ga$ (NM) at 200 GPa, (b) *P-62m*-$U_2Ga_3$ (NM) at 200 GPa, (c) *P6/mmm*-$UGa_2$ (FM) at 0 GPa, (d) *Pm-3m*-$UGa_3$ (AFM) at 0 GPa, (e) *P6/mmm*-$UGa_3$ (NM) at 50 GPa, respectively.

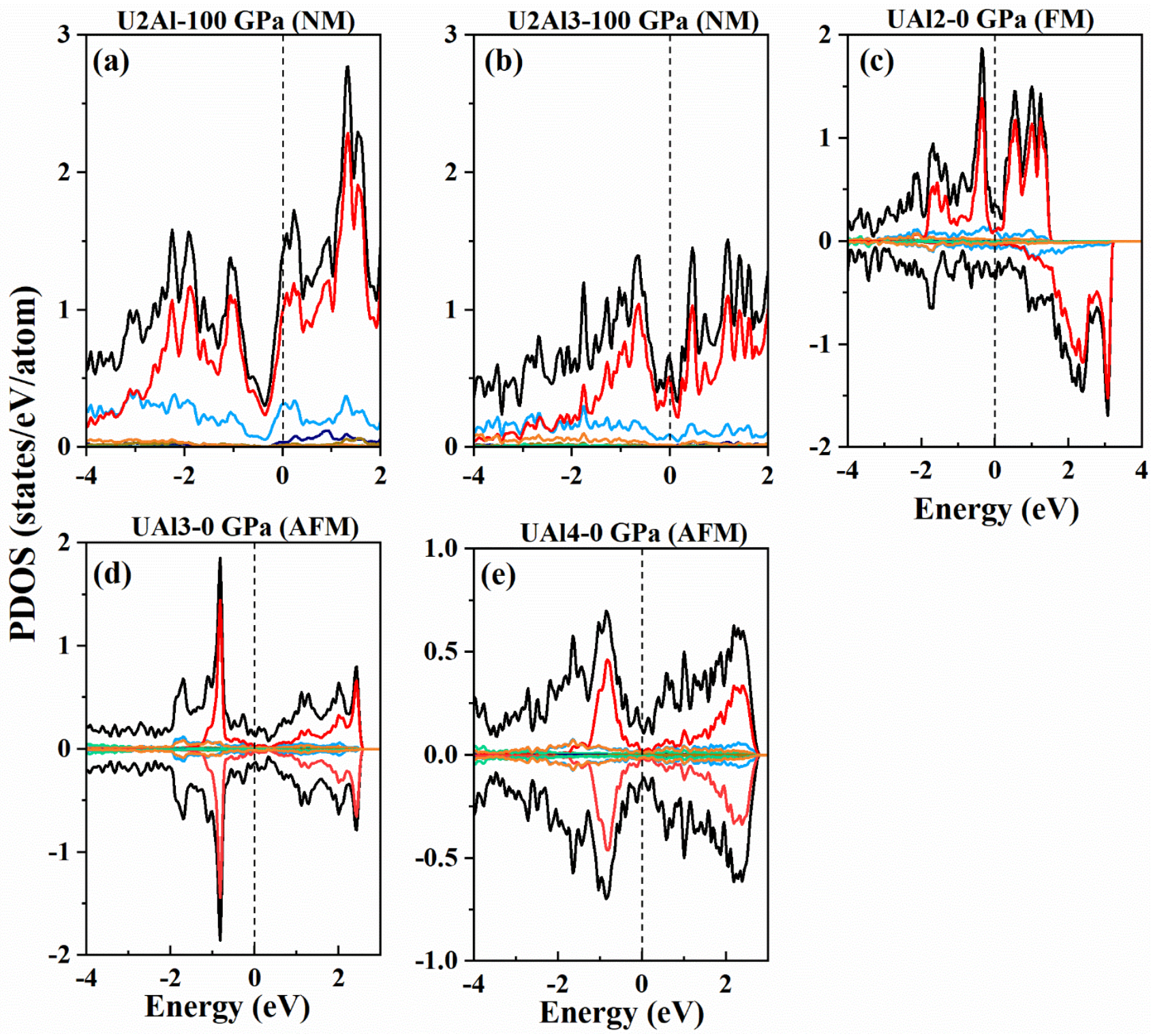

**Fig. S12** (color online) Calculated electronic projected density of states (PDOS) for (a) *P6/mmm*-$U_2Al$ (NM) at 100 GPa, (b) *P-62m*-$U_2Al_3$ (NM) at 100 GPa, (c) *P6₃/mmc*-$UAl_2$ (FM) at 0 GPa, (d) *Pm-3m*-$UAl_3$ (AFM) at 0 GPa, (e) *Imma*-$UAl_4$ (AFM) at 0 GPa, respectively.

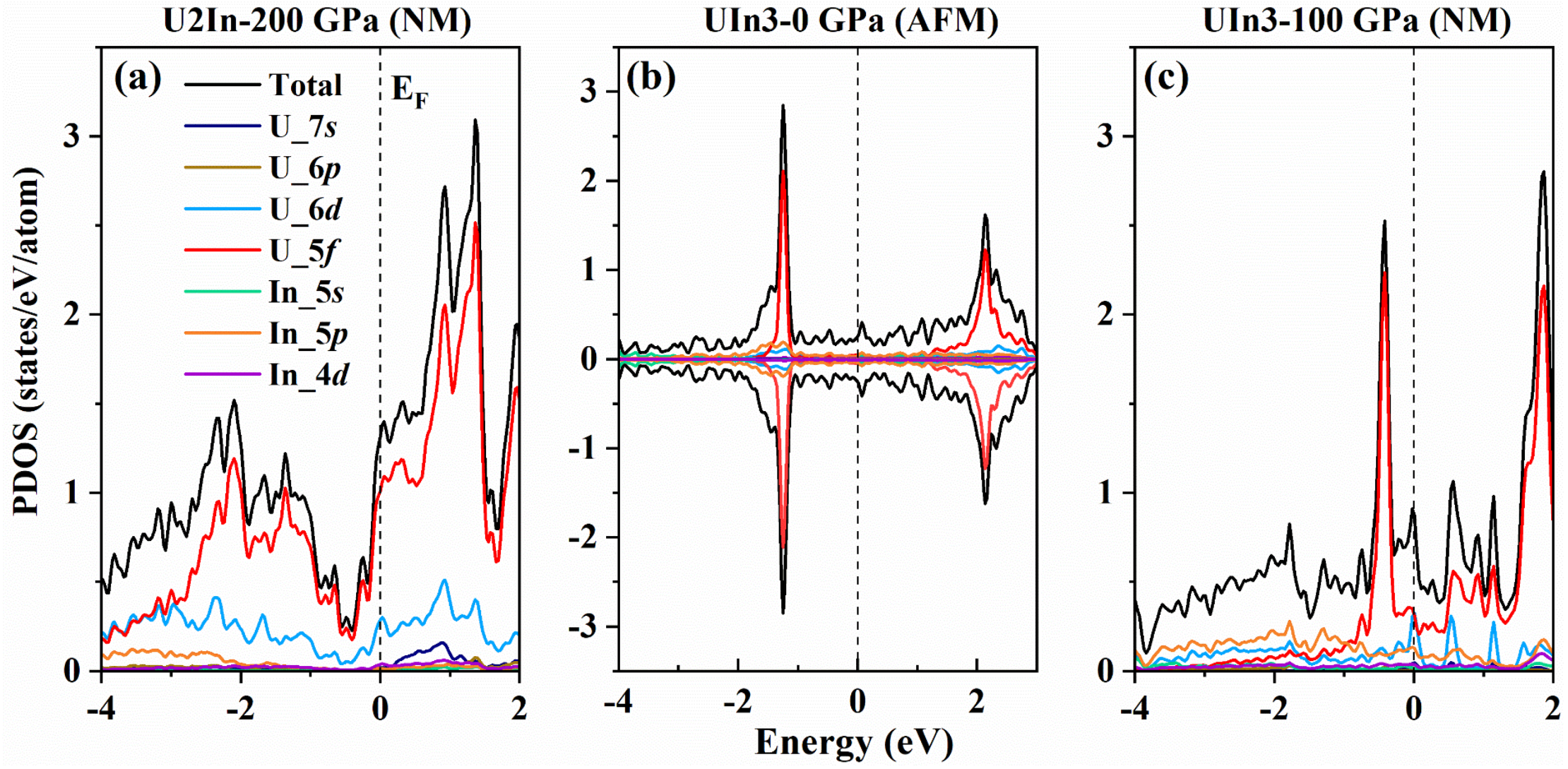

**Fig. S13** (color online) Calculated electronic projected density of states (PDOS) for (a) *P6/mmm*-$U_2In$ (NM) at 200 GPa, (b) *Pm-3m*-$UIn_3$ (AFM) at 0 GPa, (c) *P6/mmm*-$UIn_3$ (NM) at 100 GPa, respectively.

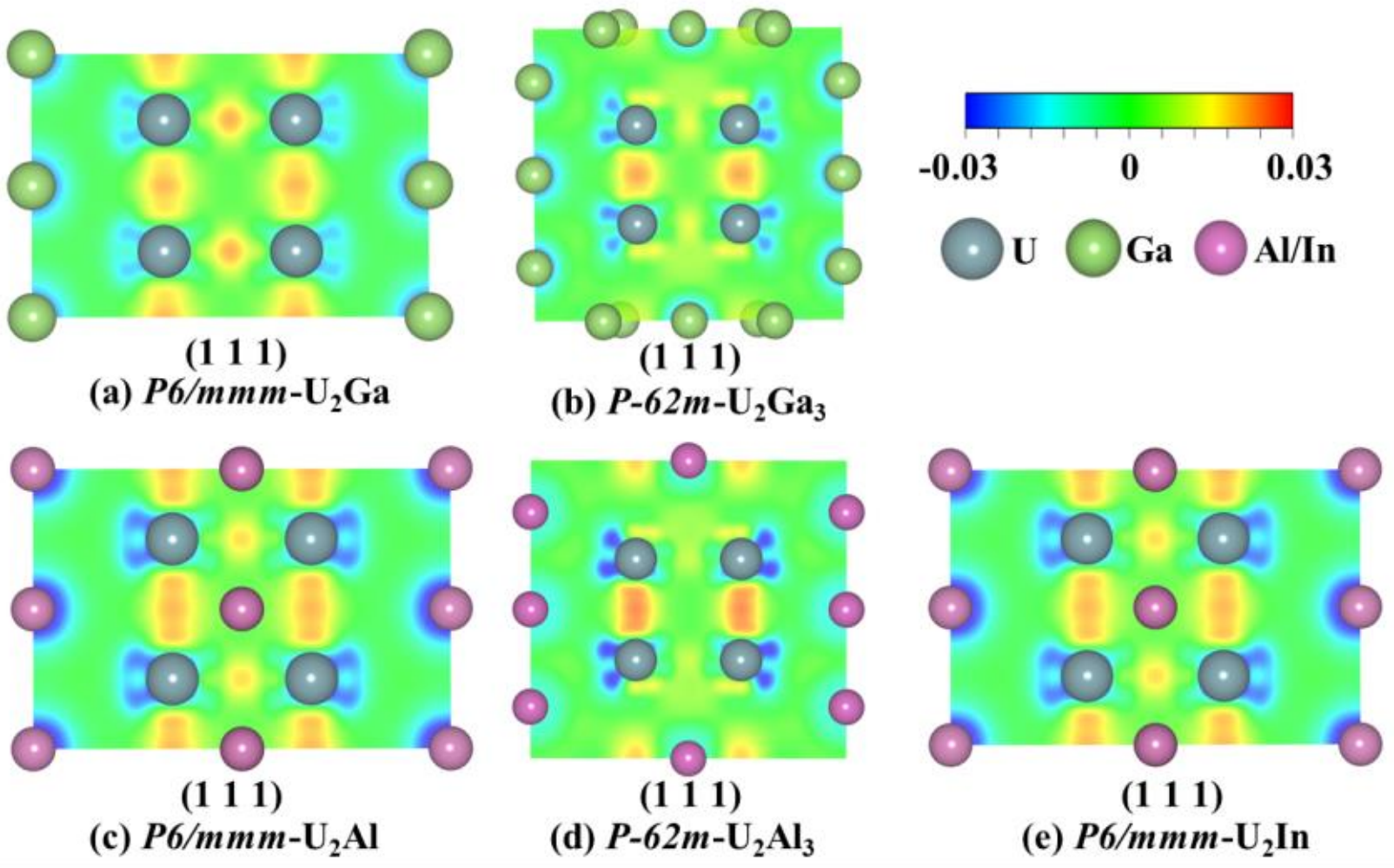

**Fig. S14** (color online) Differential charge density for stable ordered phases in U-M (M=Ga, Al, In) systems: (a) *P6/mmm*-$U_2Ga$ at 150 GPa, (b) *P-62m*-$U_2Ga_3$ at 150 GPa; (c) *P6/mmm*-$U_2Al$ at 100 GPa, (d) *P-62m*-$U_2Al_3$ at 100 GPas; (e) *P6/mmm*-$U_2In$ at 200 GPa, respectively. The blue color indicates the loss of electrons, and yellow indicates the accumulation of charges between neighbouring uranium atoms. The formation of U-U bonds is evident.

## F. The $\Delta E$-$\Delta N_U$ relationship for other compounds

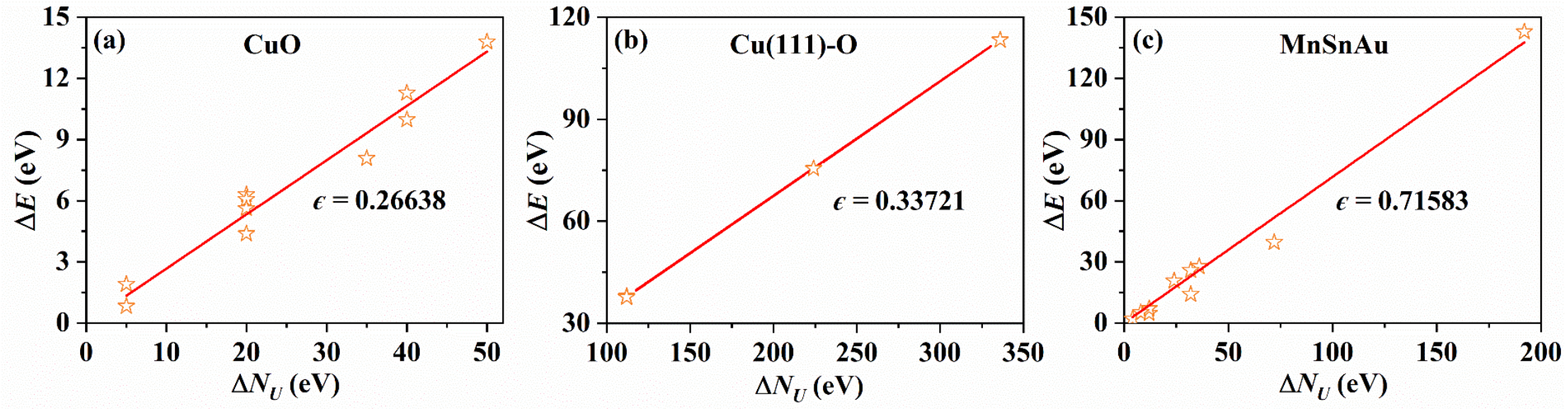

**Fig. S15** The $\Delta E$-$\Delta N_U$ relationships and the correction constants of the LCM method for the (a) binary bulk Cu-O, (b) oxygen adsorption on Cu(111) surface Cu(111)-O, and (c) ternary MnSnAu systems.

# SIII. Supplementary table

**Table S1.** Lattice parameters, atomic coordinates and Wyckoff sites occupation of the newly discovered stable phases *P6/mmm*-$U_2Ga$, *P-62m*-$U_2Ga_3$, *P6/mmm*-$UGa_3$, *P-62m*-$U_2Al_3$, *P6/mmm*-$U_2Al$, *P6/mmm*-$UIn_3$, and *P6/mmm*-$U_2In$ at the given pressure, respectively.

| Phase | Lattice parameters (Å) | Atom | Site | Atomic coordinates |
|---|---|---|---|---|
| *P6/mmm*-$U_2Ga$ (150 GPa) | a=b=4.4338 c=2.5644 α=β=90° γ=120° | U | *2d* | (0.33333, 0.66667, 0.50000) |
| | | Ga | *1a* | (0.00000, 0.00000, 0.00000) |
| *P-62m*-$U_2Ga_3$ (150 GPa) | a=b=4.3376 c=7.0907 α=β=90° γ=120° | U | *4h* | (0.33333, 0.66667, 0.33021) |
| | | Ga | *1b*<br>*2e*<br>*3f* | (0.00000, 0.00000, 0.50000)<br>(0.00000, 0.00000, 0.18315)<br>(0.55872, 0.00000, 0.00000) |
| *P6/mmm*-$UGa_3$ (50 GPa) | a=b=5.11031 c=2.5183 α=β=90° γ=120° | U | *1a* | (0.00000, 0.00000, 0.00000) |
| | | Ga | *1a*<br>*1a*<br>*1a* | (0.50000, 0.50000, 0.50000)<br>(0.00000, 0.50000, 0.50000)<br>(0.50000, 0.00000, 0.50000) |
| *P-62m*-$U_2Al_3$ (100 GPa) | a=b=4.4366 c=7.2464 α=β=90° γ=120° | U | *4h* | (0.33333, 0.66667, 0.33032) |
| | | Al | *3f*<br>*2e*<br>*1b* | (0.50030, 0.00000, 0.00000)<br>(0.00000, 0.00000, 0.17293)<br>(0.00000, 0.00000, 0.50000) |
| *P6/mmm*-$U_2Al$ (100 GPa) | a=b=4.3063 c=2.4901 α=β=90° γ=120° | U | *2d* | (0.33333, 0.66667, 0.50000) |
| | | Al | *1a* | (0.00000, 0.00000, 0.00000) |
| *P6/mmm*-$UIn_3$ (100 GPa) | a=b=5.2812 c=2.5592 α=β=90° γ=120° | U | *1a* | (0.00000, 0.00000, 0.00000) |
| | | In | *1a*<br>*1a*<br>*1a* | (0.50000, 0.50000, 0.50000)<br>(0.00000, 0.50000, 0.50000)<br>(0.50000, 0.00000, 0.50000) |
| *P6/mmm*-$U_2In$ (200 GPa) | a=b=4.2377 c=2.4499 α=β=90° γ=120° | U | *2d* | (0.33333, 0.66667, 0.50000) |
| | | In | *1a* | (0.00000, 0.00000, 0.00000) |

# Supplementary references